# Scalable synthesis of 2D van der Waals superlattices

Michael J. Motala,[1,2+] Xiang Zhang,[3+] Pawan Kumar,[4,10] Eliezer F. Oliveira,[5,6] Anna Benton,[1,2,7] Paige Miesle,[1,2,7] Rahul Rao,[1] Peter R. Stevenson,[1] David Moore,[1,2] Adam Alfieri,[4] Jason Lynch,[4] Guanhui Gao,[3] Sijie Ma,[3] Hanyu Zhu,[3] Zhe Wang,[8] Ivan Petrov,[9] Eric A. Stach,[10] W. Joshua Kennedy,[1] Shiva Vengala,[11] James M. Tour,[3,8] Douglas S. Galvao,[5,6] Deep Jariwala,[4] Christopher Muratore,[7] Michael Snure,[11]* Pulickel M. Ajayan,[3]* Nicholas R. Glavin[1]*

[1]Materials and Manufacturing Directorate, Air Force Research Laboratory, Wright-Patterson AFB, OH 45433, USA
[2]UES Inc., Beavercreek, OH 45431, USA
[3]Materials Science and Nano Engineering, Rice University, Houston, TX 77005, USA
[4]Electrical and Systems Engineering, University of Pennsylvania, Philadelphia, PA 19104, USA
[5]Group of Organic Solids and New Materials, Gleb Wataghin Institute of Physics, University of Campinas (UNICAMP), Campinas, SP, Brazil
[6]Center for Computational Engineering & Sciences (CCES), University of Campinas (UNICAMP), Campinas, SP, Brazil
[7]Department of Chemical and Materials Engineering, University of Dayton, Dayton, OH, 45469, USA
[8]Department of Chemistry, Rice University, Houston, TX 77005, USA
[9]Department of Materials Science, Physics, and the Frederick Seitz Materials Research Laboratory, University of Illinois, Urbana, IL 61801, USA
[10]Materials Science and Engineering, University of Pennsylvania, Philadelphia, PA 19104, USA
[11]Sensors Directorate, Air Force Research Laboratory, Wright-Patterson AFB, OH 45433, USA

[+]These authors contributed equally to first authorship
*Co-corresponding authors: michael.snure.1@us.af.mil, ajayan@rice.edu, nicholas.glavin.1@us.af.mil

**Abstract**

Heterostructure materials form the basis of much of modern electronics, from transistors to lasers and light-emitting diodes. Recent years have seen a renewed focus on creating heterostructures through the vertical integration of two-dimensional materials, including graphene, hexagonal boron nitride, and transition metal dichalcogenides (TMDCs). However, fundamental challenges associated with materials processing have limited material quality and impeded scalability. We demonstrate a method to convert sub-nanometer metal films deposited on silicon and sapphire into TMDC heterostructures through vapor-phase processing. The resulting heterostructures and superlattices exhibit novel properties compared with stand-alone TMDCs, including reduced bandgap, enhanced light-matter coupling, and improved catalytic performance. This robust and scalable synthetic method provides new opportunities to generate a wide range of artificially stacked 2D superlattices with controlled morphology and composition.

**Main Text**

Atomic layer-by-layer assembly of two-dimensional (2D) heterostructures has emerged as a versatile platform to realize multifunctional materials by design.[1–4] Over two decades of work has shown that 2D heterostructures generated using layer-by-layer assembly can have properties that are both unique and distinct from their individual components. Prominent examples include the observation of superconductivity in twisted graphene[5–7] and valley-polarized carrier excitations in transition metal dichalcogenide (TMDC) heterostructures.[8] The overwhelming majority of these heterostructures are meticulouslyMO constructed by exfoliation and stacking small flakes of material one layer at a time, reducing the applicability for mass-production. Recent advances in metal-organic chemical vapor deposition (MOCVD) of 2D heterostructures and superlattices show promise in allowing for large-area films, but the slow growth rate (monolayer lateral growth rate of ~ 0.15 nm/min per layer for superlattices) and transfer requirements present significant microfabrication challenges.[8] It is therefore paramount that future systems of 2D superlattices rely on strategies that are performed not just at wafer scale, but with feasible processing times in-line with conventional device fabrication. Furthermore, it will be critical that films are grown directly on support wafers without the requirement of post-growth transfer and with properties that are easily tailored to the application of interest.

In this work, we demonstrate a scalable route towards all-semiconducting 2D TMDC van der Waals superlattices through selenization (or sulfurization) of alternating stacks of sub-nm metal films. Uniform deposition of molybdenum (Mo) and tungsten (W) films with thickness as low as 0.7 nm were realized through kinetically-controlled magnetron sputtering. The use of metal ion energies approaching 60 eV resulted in continuous, uniform metal films with low surface roughness (<0.2 nm RMS). The conversion to superlattice structures was accomplished by thermally reacting the wafer-scale ultrathin metal heterostructures in either hydrogen selenide ($H_2Se$) or hydrogen sulfide ($H_2S$). This led to the formation of few-layer $MoSe_2$/$WSe_2$ or $MoS_2$/$WS_2$ alternating layers, with each TMDC layer being four to five atomic layers in thickness. The entire processing time for superlattice film fabrication on either sapphire or $SiO_2$ was approximately 30 minutes, with a metal growth rate of 0.2 nm/second followed by annealing in a chalcogen hydride ($H_2Se$ or $H_2S$) between 400 and 800 °C.

**Wafer-scale conversion of sub-nm metal films to TMDCs**

Reacting metal thin films with chalcogen vapor at elevated temperatures, as depicted in Figure 1a, has been an established means of 2D TMDC synthesis.[9–11] For processing at scale (> cm$^2$) and in few-layer configurations (<5 monolayers), however, 2D materials grown via this method are generally discontinuous and demonstrate non-uniform properties. To encourage a transition from 3D island formation to 2D layer-by-layer growth of the metal films, the power applied to the metal target source during sputtering was modulated to adjust the kinetic energy distributions of ionized species. The resultant energy distributions measured for Mo ions is shown in Figure 1b, depicting a maximum ion energy approaching 60 eV for pulsed DC sputtering.[12] The dependence of electrical conductivity on metal film thickness (Figure 1b, inset) demonstrates that the higher kinetic energy ions in pulsed direct current (DC) power modulation resulted in 2D continuity for the thinnest films, approaching five atomic layers (further details in supplementary information Figures S1-4). The ultrathin metal films produced by this technique are contiguous over large areas, with an atomically smooth surface, as shown by the AFM image in Figure 1c and the optical image (inset, Figure 1c) of the 1 cm$^2$ sapphire substrate coated by 0.7 nm Mo. Similar results were confined for ultrathin W films. These metal films were then selenized at 500 Torr under a flow of $H_2Se$ (150 sccm), $N_2$ (160 sccm) and $H_2$ (10 sccm) over a range of temperatures from 400-800 °C (Figure 1d) to convert to the respective TMDC crystalline structure. Here $H_2Se$ is preferred over Se powder sources due to its low decomposition temperature and simple pyrolysis products.[13–15] The measured Raman spectra and optical constants (i.e., the refractice index, $n$, and extinction coefficient, $k$) for monolithic five-layer $MoSe_2$ (Figure 1e-f) and $WSe_2$ (Figure 1g-h) films synthesized using this two-step method reveal properties comparable to films grown via MOCVD. As MOCVD is considered to be the state of the art in scalable 2D TMDC synthesis, the two-step growth methodology described here represents a viable alternative to uniform growth of multilayer 2D materials.[16,17]

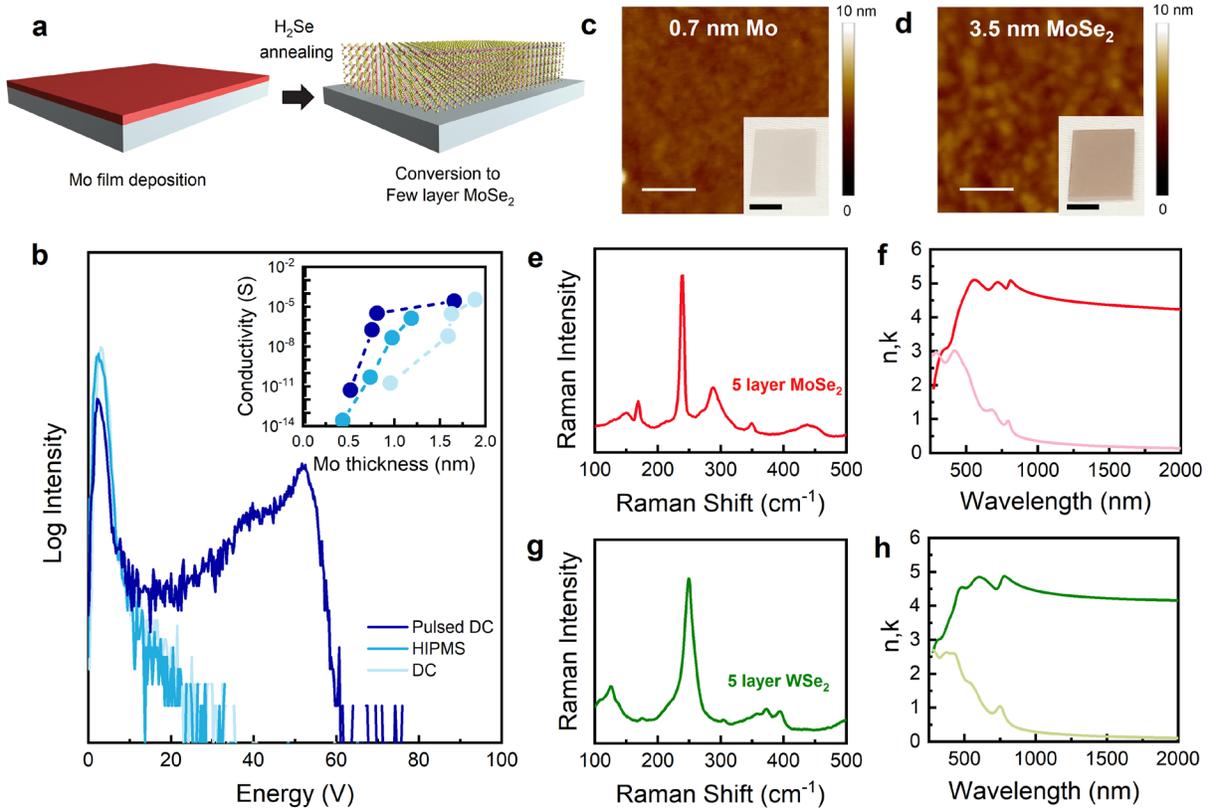

*Figure 1: Two-step synthesis of few-layer TMDCs. a) Schematic of conversion of Mo or W thin films to subsequent $MoSe_2$ or $WSe_2$ using $H_2Se$ annealing, b) $Mo^+$ Ion energy distribution for various sputtering techniques and inset describes film conductivity as a function of thickness of Mo, c) AFM height image indicating 0.7 nm Mo film and d) 3.5 nm $MoSe_2$ film with 50 nm scalebars on AFM image equivalent and inset images of wafers with 0.5 cm scalebars, e) Raman and f) n (darker color) and k (lighter color) values of five layer $MoSe_2$, g) Raman and h) n (darker color) and k (lighter color) values of five layer $WSe_2$.*

Sequential sputtering of Mo and W metal sources was employed for synthesis of metal heterostructure precursor films composed of alternating five layer thick films of Mo and W as shown in Figure 2a. The presence of alternating metal layers was confirmed with cross-sectional energy dispersive x-ray spectroscopy (EDS). The red and green colored regions in the cross-sectional EDS map in Figure 2b clearly show discrete horizontal Mo and W layers with little metal film intermixing. For simplicity, we have defined the number of repeating units of Mo/W as N, where N=2 represents a film composed of Mo/W/Mo/W structure. Following metal deposition, the stacked metal films were selenized similar to the monolithic TMDC films to produce the superlattices. The opacity of the monolithic metal precursor films and superlattice stacks up to N=8 increased (Figure 2c) demonstrating uniformity of the films throughout the thickness range in this study. Figure 2d depicts AFM-measured thickness and roughness of selenized metal films after annealing at 600 °C, with insets showing a slight increase in the selenized N = 1 film roughness compared to the metal precursor stack. The roughness values increased significantly beyond N=4, suggesting a transition from atomic-scale roughness (<0.5 nm) for N=0, 1, and 2 films to nanoscale roughness (>1.0 nm) for films with N ≥ 4. These trends indicate a potential morphology change in the film structure at a certain thickness, and are explored further below. Thickness-dependent Raman spectra from the superlattices annealed at 600°C (Figure 2e) exhibit peaks corresponding to both $MoSe_2$ and $WSe_2$, which are resolved by fitting the spectra to Lorentzian peaks. Peaks from $MoSe_2$ and $WSe_2$ were observed on both sapphire and $SiO_2$ substrates (further details in supplementary information Figure S14-15) and provide further proof of distinct TMDC layers within the structure. With increasing thickness, the $A_{1g}$ peak frequencies from $MoSe_2$ ($WSe_2$) blueshift (redshift) in frequency, merging into a single peak at N = 16 (Figure 2f). These peak shifts suggest increasing strain and alloying with thickness, an effect that has been observed previously in TMDC heterostructures.[18] As mentioned above, one significant advantage of the two-step synthesis technique is the large area uniformity, as indicated by the $A_{1g}$ Raman intensity map of a 2-inch wafer of an N=2 film and Raman spectra collected from 10 random spots on the film (Figure 2g and h, respectively).

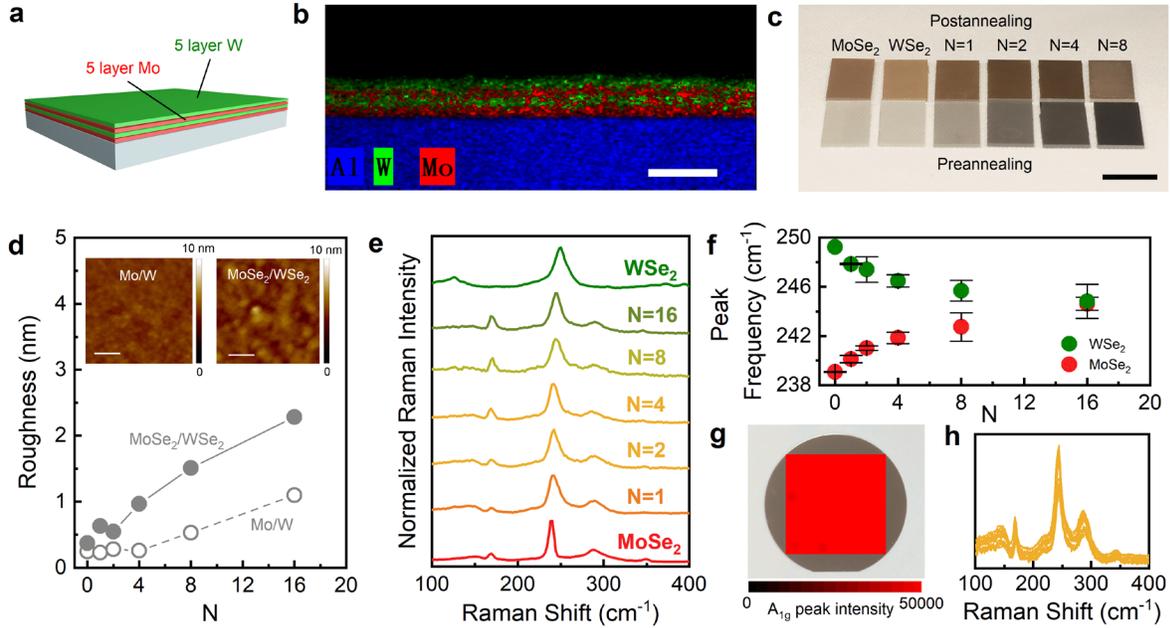

*Figure 2: Synthesis of MoSe$_2$/WSe$_2$ superlattices. a) schematic of Mo/W superlattice structure, b) EDS map of Mo/W/Mo/W heterostructure with scalebar equivalent to 4 nm, c) optical image wafer-scale few-layer films and superlattices on sapphire substrates with scalebar equivalent to 1 cm, d) AFM surface roughness values with inset depicting atomically-flat films and heterostructures with scalebar equivalent to 50 nm, e) thickness-dependent Raman spectra, f) Raman peak shift with increasing N value, and g) Raman map from a 2-inch wafer showing A$_{1g}$ peak intensity, and h) Raman spectra collected from 10 random spots on the sample indicating highly uniform film properties.*

**Modular orientation and chemistry in vdW heterostructure superlattice films**

The influence of the annealing temperature strongly influences both the orientation and chemistry of the vdW heterostructure superlattice structure as depicted schematically in Figure 3a. The full width at half maximum intensity (FWHM, $\Gamma_{A1g}$) of the A$_{1g}$ peak from MoSe$_2$ (20 cm$^{-1}$) is found to be mostly insensitive to the selenization temperature up to 600 °C and reduces sharply for the film annealed at 800 °C (9 cm$^{-1}$). This decrease in $\Gamma_{A1g}$ at high temperatures can be attributed to improved crystallinity with annealing temperature. Furthermore, significant improvements in crystallinity and grain size are observed in scanning transmission electron microscope (STEM) cross-sections (Figure 3c-e) indicating grain size evolution from 20 nm to 60 nm in the annealed films from 400°C to 800°C (further details in supplementary information section S4). Moreover, the orientation of annealed films was shown to be highly temperature dependent, as confirmed by the polarized Raman analysis shown in Figure 3b. As the A$_{1g}$/E$_{2g}$ intensity ratio (I$_{A1g}$/I$_{E2g}$) can be used as an indicator of layer orientation under the cross-polarized configuration, a transition from a mixed/random grain orientation to a more parallel layered orientation was observed with increasing annealing temperature.[20,21] This orientation effect was confirmed via cross-sectional STEM images in Figure 3c-e (further details in supporting information Figure S14-15).

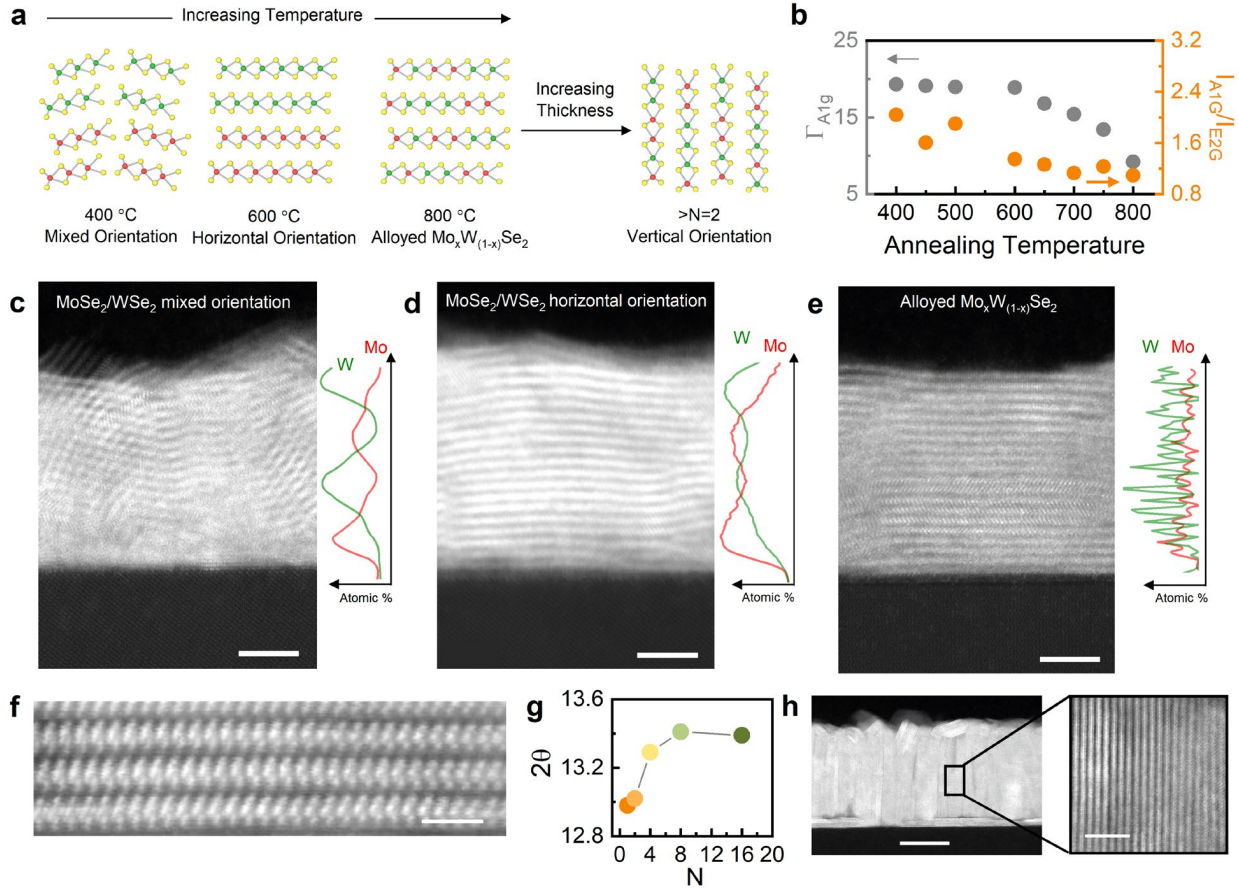

*Figure 3: Temperature and thickness-dependent orientation control in superlattices. a) Schematic of the role of temperature and thickness in superlattice chemistry and orientation, b) Raman $A_{1g}$ FWHM and polarized Raman as a function of annealing temperature, c) STEM and EDS spectra of N=2 sample annealed at 400 °C, d) 600 °C and e) 800 °C, with scalebar equivalent to 2 nm, f) atomically resolved HAADF STEM images depicting TMDC structure with scalebar equivalent to 1 nm, g) XRD peak location shift with increasing N value, and h) TEM and atomically resolved STEM images of vertically oriented TMDC alloys with scalebar equivalent to 25 nm and 3 nm, respectively.*

The improved layer crystallinity at higher temperatures comes as the consequence of a transition from discrete superlattices to alloying in N=2 films (Figure 3c-e). The cross-sectional EDS line profile from N=2 films annealed at 400 and 600 ºC showed the Mo/W metal heterostructures were successfully converted to MoSe$_2$/WSe$_2$ superlattices with distinct layering of Mo (W) rich W (Mo) poor regions with an N=2 repeat unit. In the 800 °C annealed films, however, Mo and W concentrations indicated that rather than forming discrete $MoSe_2/WSe_2$ layers, the layered material instead forms a $Mo_{1-x}W_xSe_2$ alloy. With increasing temperature the entropic contribution to Gibbs free energy increases favoring mixing and the formation of alloys in TMDCs.[22,23] This establishes a temperature limit for forming TMDC superlattices and not alloys at temperatures at or below 600 °C. Considering the thermal competition between high temperatures required for highly crystalline materials and lower temperatures for retaining distinct $MoSe_2$/ $WSe_2$ layers, a conversion temperature of 600 °C was optimum temperature for retaining chemical separation and crystal quality, as observed in the atomically resolved TEM image in Figure 3f. Additionally, the formation of a $MoS_2/WS_2$ superlattices was confirmed using similar conditions as that for $MoSe_2/WSe_2$ superlattices (further details are found in supplementary information Figure S23).

Analysis of X-ray diffraction (XRD) spectra shown in Figure 3g depicted commensurate trends in peak shift with N, indicative of expansion of both TMDC lattices in the z direction for films with N > 2. As the superlattice layer thickness increased there was a profound effect on layer orientation as observed in Figure 3h. For N=8 superlattices, layers became primarily vertical folding and bending at the top and bottom surfaces (Figure 3h) consistent with previous reports showing the density of vertically grown layers increased with metal or metal oxide film thickness.[24,25] As the Mo/W superlattice layers selenize the films undergo lattice expansion, which induced a significant amount of strain in thicker films. By reorienting vertically, the strain was relaxed through vertical expansion.[24,26] Additionally, the different diffusion rates of dichalcogenide atoms through the van der Waals interlayer gaps and across the layers was a critical factor for the thickness dependent growth orientation.[25–28]

**Emergent optical and catalytic properties**

The wafer-scale superlattice structures represent an enticing platform for designing bulk optical properties through layering of 2D TMDCs of controllable orientation and chemistry. The imaginary part $\varepsilon_2$ of the complex dielectric function of $MoSe_2$, $WSe_2$, N=1 and N=2 superlattice, N=2 and N=8 alloy are shown in Figure 4a (additional details in supplementary information Figure S28). The spectra of $MoSe_2$ and WSe$_2$ agree well with previous reports, where two excitonic characteristic peaks are observed at 1.6 eV (A) and 1.8 eV (B) for MoSe$_2$, 1.7 eV (A) and 2.1 eV (B) for WSe$_2$, respectively, which originate from the spin-split direct gap transitions at the K point of the Brillouin zone.[29–31] Several unique critical points (CPs) were observed in the $\varepsilon_2$ spectra of N=1 and N=2 superlattice as labeled by A-D in Figure 4a, which differ in location and relative intensity from the individual MoSe$_2$ and WSe$_2$ spectra. These CPs located at 1.4 eV (A), 1.7 eV (B), 2.2 eV (C) and 2.6 eV (D) matched well with the values predicted by density functional theory (DFT) and random phase approximation (RPA) displayed in Figure 4b. The broad FWHM of these experimental CPs can be attributed to multifarious and competitive optical transitions and carrier relaxation channels.[32] These expected transitions have been labeled within the electronic band structure in Figure 4c (further details in supplementary information Figure S24). Similar CPs were observed in the $\varepsilon_2$ spectra of N=2 alloy film, but the amplitude was lower than N=1 and N=2 superlattice, indicating that with the same thickness, the optical response of the superlattice film was higher than the alloy film. Additionally, the large refractive index above 4.9 (further details in supplementary information Figure S28) for the N=2 superlattice indicated that the films produced herein are excellent candidates for light trapping elements in optoelectronics. Futhermore, the highest extinction coefficient k was observed near the visible spectrum below 670 nm and was attributed to the significantly enhanced excitonic behavior induced by the quantum confinement effect. In contrast, the CPs of N=8 alloy film showed much broader FWHMs and weaker intensities compared to N=2 film but higher absorption near the infrared regime.

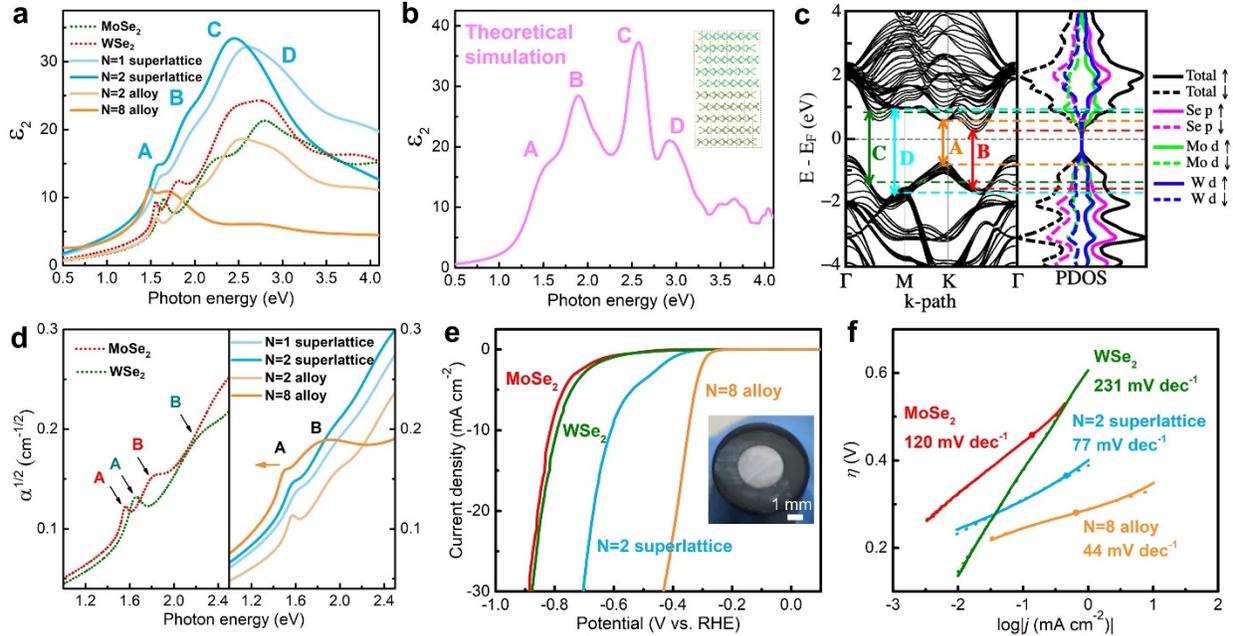

*Figure 4. Optical and catalytic performance of superlattice thin films. (a) The imaginary part ε₂ of the complex dielectric function of MoSe₂, WSe₂, N=1 and N=2 superlattice, N=2 and N=8 alloy. Critical points A to D is labeled for N=1 and N=2 superlattice, (b) Theoretically simulated ε₂ spectra of the superlattice with similar critical points A to D. The inset shows the atomic structure of one period of the superlattice, (c) The electronic band structure and the projected density of states of superlattice film, (d) The absorption spectra α$^{1/2}$ of MoSe₂, WSe₂, N=1 and N=2 superlattice, N=2 and N=8 alloy with exciton peaks A and B, (e) Typical cathodic polarization curves and (f) corresponding Tafel plots of MoSe₂, WSe₂, N=2 superlattice, N=8 alloy. The inset in (e) shows a picture of transferred film on a 3 mm diameter glassy carbon electrode.*

The absorption spectra in Figure 4d depict two expected exciton peaks A and B for both MoSe₂ (1.57 eV and 1.82 eV) and WSe₂ (1.66 eV and 2.22 eV).[33,34] In both the superlattice and alloy films, the A exciton energy decreased with the film thickness, with peak position at 1.59 eV for N=1 superlattice film, 1.57 eV for N=2 superlattice and alloy films, and 1.49 eV for N=8 vertically-oriented alloy film. Using the Tauc plot estimation, the optical bandgap of these films were estimated to be 1.48 eV for MoSe₂, 1.53 eV for WSe₂, 1.01 eV for N=1 superlattice film, 1.04 eV for N=2 superlattice film, 1.22 eV for N=2 alloy film, 1.08 for N=8 alloy film. The decrease in bandgap in the superlatice films is supported in the simulated electronic band structure and the projected density of states (PDOS), which predicted that the N=2 superlattice film has an indirect bandgap of 0.722 eV. This bandgap emerges at the transition from the Γ point to a K point approximately in the middle of the K-Γ path. The simulated electronic bandgap was about 300 meV lower than the optical bandgap extracted from the experimental results, which can be attributed to a common artifact of predicting bandgaps by DFT and the deviation of the Tauc method.[35–37] The square of the wave function of the valence band maximum (VBM) and the conducting band minimum (CBM) (further details shown in supplementary information Figure S24-27) suggesds that VMB and CBM are localized in $WSe_2$ layers and $MoSe_2$ layers, respectively, in agreement with the DFT simulations. The electrostatic interactions and the intralayer spin orbit coupling between $WSe_2$ and $MoSe_2$ layers in the superlattice leads to bands splitting and bandgap decreasing,[38,39] so that even few-layer $WSe_2$/few-layer $MoSe_2$ struture exhibited a narrower bandgap than monolayer $WSe_2$/monolayer $MoSe_2$. The unique electronic band structure of the superlattices supports the findings that the layer-by-layer assembly of 2D nanomaterials reveals an exciting platform for designing both electronic and optical properties that differ from the properties of the individual layered constitutuents.

Finally, the ability to control both orientation and chemistry in the superlattice and alloyed structures provided a unique platform for new catalytic properties and applications. To evaluate the HER catalytic activity of $MoSe_2$, $WSe_2$, $MoSe_2$/$WSe_2$ N=2 superlattice, and N=8 alloy, the samples were transferred from the sapphire substrates to glassy carbon electrodes using a PMMA-assisted wet transfer method as shown in Figure 4e. Typical cathodic polarization curves and corresponding Tafel plots are shown in Figure 4e and 4f, respectively, where it was observed that N=2 superlattice exhibited a better catalytic performance compared to monolayer $MoSe_2$ or $WSe_2$. This was primarily attributed to the the narrower band gap in the superlattice structure as compared to that of $MoSe_2$ or $WSe_2$. Additionally, the N=8 alloy showed the smallest onset overpotential and Tafel slope among all the films, which can be attributed to maximal edge sites on the vertically oriented films. It is worth noting the observed 44 mV/dec Tafel slope of N=8 alloy film was among the lowest reported Tafel slopes of transition metal diselenide films.[40–43]

**Conclusions**

The synthesis approach for 2D van der Waals superlattices using conversion of sequentially layered metal heterostructures represents a platform for layer-by-layer design and assembly of 2D semiconductors in a manner that is highly scalable. The scalability in this case was not only due to large area, but the short processing time (~ 30 minutes), the reasonable temperatures (400-800 °C), and low cost substrates ($SiO_2$ or sapphire). The production of 2 inch wafers was easily achieved and limited only through the size of the equipment available, as sputtering and annealing are both industry-scalable processes. Finally, the material set described here represents an intruiging path forward to tune bulk properties of materials through careful layer-by-layer engineering of nanomaterials.

**Methods**

*Synthesis*

A custom-built magentron sputtering system was employed for synthesis of ultrathin metal films (minimum base pressure of $3x10^{-8}$ Torr) and contained two 1.3" Mac sputter sources (MeiVasc, Inc) held at a distance of 8 cm and a 30° angle from the samples. The metal TMDC precursor layers were deposited utilizing a Pinnacle Plus pulsed power supply (Advanced Energy) utilizing high purity Mo and W targets Angstrom Sciences, Inc.). The deposition process was optimized to occur at a pressure of 10 Torr under argon with a flow rate of 25 sccm. The pulsed power supplies were operated at 90W with a pulse frequency of 65 kHz and positive pulse width of 0.4 microseconds.

Subsequent to metal deposition the samples were unloaded and transferred to a vacuum tube furnace for reaction with $H_2Se$ gas. In a typical TMDC conversion the samples were pumped an hour followed by an additional hour of full pumping under hydrogen gas flow at 200 sccm. The pressure was then increased to 500 Torr and the gas adjusted to 10 sccm hydrogen, 160 sccm nitrogen. A 20 minute temperature ramp was used to heat the furnace to desired temperature. After 5 minutes at equilibrium the $H_2Se$ was turned on at a flow rate of 150 sccm. This held for thirty minutes before the chamber rapidly cooled by opening the tube furnace lid. $H_2Se$ gas was turned off once the temperature was below 400°C. The tube furnace was only brought to atmospheric pressure after completely cooling. All samples were stored in a glove box to minimize oxidation.

*Raman Spectroscopy*

Normal and polarized Raman spectra were obtained at room temperature with a Renishaw InVia microscope using a 514.5 nm wavelength excitation laser with ∼600 nm spot size (100× objective lens). The laser power was kept to <1 mW to avoid sample heating. The peak frequencies and widths were obtained by spline baseline subtraction followed by Lorentzian lineshape fitting.

*Variable Angle Spectroscopic Ellipsometry*

A J.A. Woollam RC2 Ellipsometer was used to characterize the optical properties of $MoSe_2$, $WSe_2$, and superlattice films. Optical dispersion data were collected from 300-2500 nm at 50-80° angles with 5° intervals. A Lorentz multi-oscillator formalism was used for the optical dispersion data analysis using CompleteEASE v6.55. Each respective model incorporated the Si/SiO2 or single-side polished C-plane sapphire substrate. Model parameters for each film are provided in the Supplementary Information (Table S3) and were used to determine n, k, ε1, ε2 and α for the TMDC film responses reported. For simplicity, the derived optical properties from ellipsometry represent an ensemble modeled response (meaning no modeling of individual layers within the respective stacks). However, N=1 and N=2 heterostructure films were capable of being modeled in this layer-by-layer fashion yielding overall optical responses commensurate with the non-layer-by-layer modeling approach.

*Theoretical simulations*

The DFT simulations were performed using a projector augmented wave potential (PAW) in a generalized gradient approximation (GGA) with the Perdew, Burke, and Ernzerhof (PBE) exchange-correlation functional.[44] The simulated unit cells of superlattice film were composed of 30 atoms. For the geometry optimizations, a Monkhorst-Pack k-mesh of 6x6x1 was used to sample the Brillouin zone and the Kohn−Sham orbitals were expanded in a plane-wave basis set with a kinetic energy cutoff of 60 Ry (∼ 816 eV). The k-mesh was doubled for the electronic structure simulation. The convergence for energy was set to be $10^{-6}$ eV between two consecutive steps, and the maximum Hellmann−Feynman forces acting on each atom was set to be less than 0.01 eV/Å upon ionic relaxation. To account for the interlayer interactions, the D3 dispersion corrections were applied for van der Waals interactions (DFT-D3).[45] The effects of spin-orbit coupling (SOC) were considered in all simulations. The dielectric function $\epsilon(\omega)$ was calculated within the Random Phase Approximation (RPA)[46] based on DFT ground-state simulations. All simulations were performed with the computational code Quantum Espresso (QE).[47] It is worth noting that the RPA/DFT simulation with QE is implemented only for norm-conservative pseudopotentials (NC). More details about theoretical simulations can be found in Supplementary Information.

*HER measurements*

HER experiments were carried out using a CHI708D electrochemical workstation (CH Instruments, Inc.) with a three-electrode configuration. A graphite rod and a $Hg/Hg_2SO_4$ ($K_2SO_4$, saturated) were used as the counter electrode and the reference electrode, respectively. Typically, a continuous Ar bubble was introduced into 0.5 M $H_2SO_4$ electrolyte to achieve saturation before the HER test. The linear sweep voltammograms were obtained with a scan rate of 5 mV/s. The $Hg/Hg_2SO_4$ ($K_2SO_4$, saturated) reference electrode was calibrated with a reversible hydrogen electrode.

**Acknowledgements**

N.G. and M.S. acknowledge the support from the Air Force Office of Scientific Research under grant number FA9550-19RYCOR050. EFO and DSG would like to thank the Brazilian agency FAPESP (Grants 2013/08293-7, 2016/18499-0, and 2019/07157-9) for financial support. Computational support from the Center for Computational Engineering and Sciences at Unicamp through the FAPESP/CEPID Grant No. 2013/08293-7 and the Center for Scientific Computing (NCC/GridUNESP) of São Paulo State University (UNESP) is also acknowledged. The work in the laboratory of JMT was funded by the Air Force Office of Scientific Research (FA9550-19-1-0296). H.Z. was supported by the U.S. National Science Foundation (NSF) under award number DMR 2005096. D.J. acknowledges primary support for this work by the Air Force Office of Scientific Research (AFOSR) FA9550-21-1-0035. D.J and A.A. also acknowledge support from the AFOSR under award number FA2386-21-1-4063. D.J. and J.L. also acknowledge support from the AFOSR under award number FA2386-20-1-4074. D.J., E.A.S. and P. K. acknowledge support from National Science Foundation (DMR-1905853) and support from University of Pennsylvania Materials Research Science and Engineering Center (MRSEC) (DMR-1720530) in addition to usage of MRSEC supported facilities. The composition mapping in electron microscopy was carried out at the Singh Center for Nanotechnology at the University of Pennsylvania which is supported by the National Science Foundation (NSF) National Nanotechnology Coordinated Infrastructure Program grant NNCI-1542153.

**Author Contributions**

M.M. and X.Z. contributed equally to the first authorship in the manuscript. M.S., P.A., and N.G. contributed equally as corresponding authors in the manuscript. M.M., A.B., C.M. and N.G. developed and characterized the metal deposition process. M.M., S.V., M.S., N.G. performed and optimized the selenization process. M.M., X.Z., P.M., R.R., P.S., A.A., D.M., J.L., I.P., H.Z., Z.W., W.J.K., D.J., M.S., P.A., and N.G. aided in Raman, XPS, structural, SHG, and optical characterization. X.Z., E.O., and D.G. performed the DFT simulations. P.K., G.G., E.S., and D.J. performed TEM imaging studies. Z.W. and J.T. performed catalysis experiments. All authors contributed to drafting the manuscript.

**Additional Information**

Supplementary information is available for this paper. The supplementary information information is divided into seven sections providing further details to the reader. Section S1 describes the deposition of metal films and metal heterostructures. Section S2 provides further details on the annealing and conversion of the metal films to the superlattice and alloyed structures. In Section S3, details of the Raman and polarized Raman are provided. Section S4 describes details regarding the structural characterization of the annealed films via XRD and SHG. Section S5 provides the reader with additional TEM, EELS, and EDS characterization. In Section S6, details as to the simulations performed in this work are provided. Finally, Section S7 reveals details of the optical characterization in the superlattice and alloyed films.

# Supplementary Information

# Scalable synthesis of 2D van der Waals superlattices

Michael J. Motala,[1,2+] Xiang Zhang,[3+] Pawan Kumar,[4,10] Eliezer F. Oliveira,[5,6] Anna Benton,[1,2,7] Paige Miesle,[1,2,7] Rahul Rao,[1] Peter Stevenson,[1] David Moore,[1,2] Adam Alfieri,[4] Jason Lynch,[4] Guanhui Gao,[3] Sijie Ma,[3] Hanyu Zhu,[3] Zhe Wang,[8] Ivan Petrov,[9] Eric A. Stach,[10] William Joshua Kennedy,[1] Shiva Vengala,[11] James M. Tour,[3,8] Douglas S. Galvao,[5,6] Deep Jariwala,[4] Christopher Muratore,[7] Michael Snure,[11]* Pulickel M. Ajayan,[3]* Nicholas R. Glavin[1]*

[1]Materials and Manufacturing Directorate, Air Force Research Laboratory, Wright-Patterson AFB, OH 45433, USA
[2]UES Inc., Beavercreek, OH 45431, USA
[3]Materials Science and Nano Engineering, Rice University, Houston, TX 77005, USA
[4]Electrical and Systems Engineering, University of Pennsylvania, Philadelphia, PA 19104, USA
[5]Group of Organic Solids and New Materials, Gleb Wataghin Institute of Physics, University of Campinas (UNICAMP), Campinas, SP, Brazil.
[6]Center for Computational Engineering & Sciences (CCES), University of Campinas (UNICAMP), Campinas, SP, Brazil
[7]Department of Chemical and Materials Engineering, University of Dayton, Dayton, OH, 45469, USA
[8]Department of Chemistry, Rice University, Houston, TX 77005, USA
[9]Department of Materials Science, Physics, and the Frederick Seitz Materials Research Laboratory, University of Illinois, Urbana, IL 61801, USA
[10]Materials Science and Engineering, University of Pennsylvania, Philadelphia, PA 19104, USA
[11]Sensors Directorate, Air Force Research Laboratory, Wright-Patterson AFB, OH 45433, USA

[+]These authors contributed equally to first authorship
*Co-corresponding authors: michael.snure.1@us.af.mil, ajayan@rice.edu, nicholas.glavin.1@us.af.mil

## Supplementary Information Table of Contents

# Section S1: Deposition of ultrathin metal films

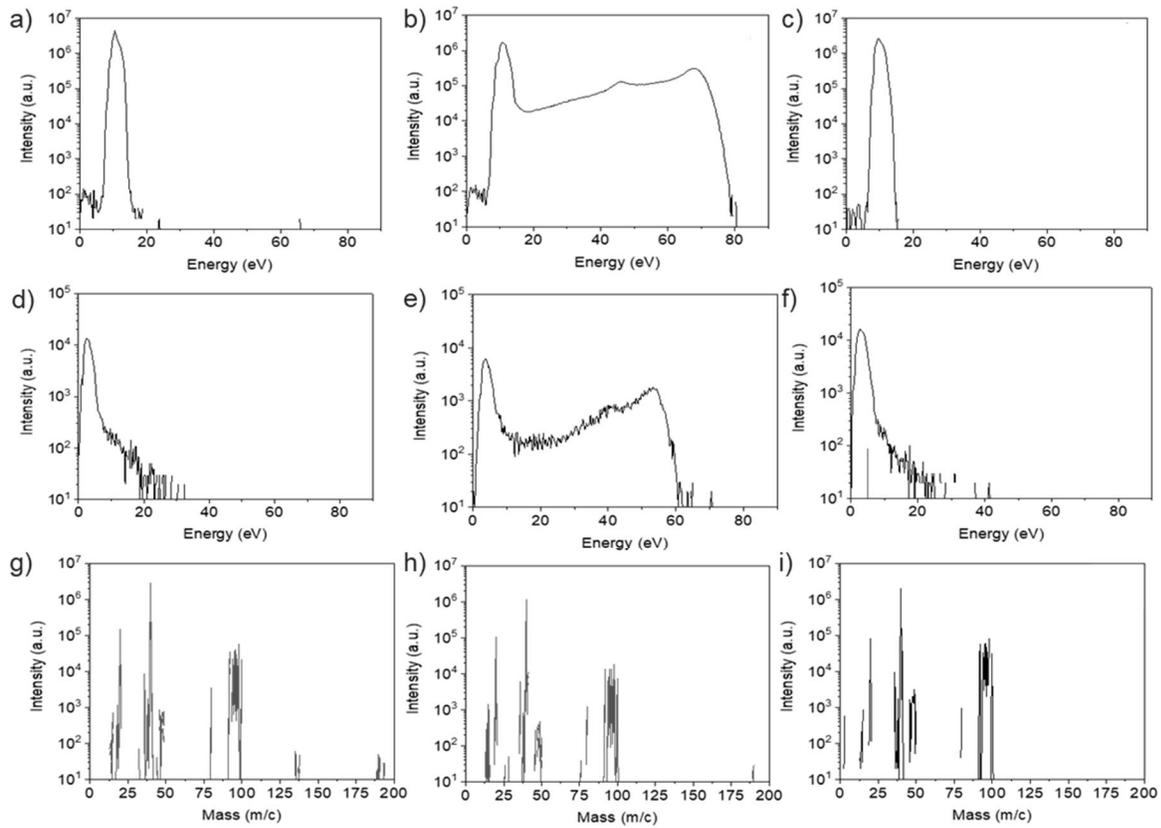

*Figure S1: Ion energy and mass distribution measurements of key species in metal film precursor processing plasmas generated with different power modulation. Figs. a-c show energy distributions measured for Ar+ ions in a) dc glow discharge, b) pulsed dc glow discharge, c) HIPIMS discharge. Figures d-f show energy distributions measured for Mo+ ions in d) dc glow discharge, e) pulsed dc glow discharge, f) HIPIMS discharge. Figs. h-j show mass distributions of all positive ions in the discharge measured at the most probable $Mo^+$ ion energy determined via analysis of a-c. $Ar^+$ (40 amu), $Ar^{++}$ (20 amu/q), and $Mo^+$ (96 amu/q, broadened by natural isotopes) comprised the majority of ionized species in the plasma.*

Ion energy and mass distributions were measured with a Hiden EQP 300 electrostatic quadrupole plasma analyzer with a sector field energy analyzer. The system was set to scan at 40 and 96 atomic mass units for energy scans, and to the most probable energy (i.e., energy correlated to global peak of energy distribution). The conditions for measurement of the ion energy and mass distributions shown in Figure S1 were identical to those used for metal precursor film growth. A table of the relevant process conditions for both plasma analysis and film depositon is included below. The increased ion energy observed when operating metal sources with the asymmetric bipolar power supply results from the natural increase of the plasma potential to rise above the most positive electrode immersed in the plasma during the positive portion of the voltage waveform delivered to the sputter target. The ion energy distributions for HIPIMS target modulation closely resemble dc ion energy distributions due to current limitations on lab-scale (33 mm diameter) sputter sources which inhibit commonly reported high energy ion formation mechanisms, however periodic interruption of growth and reduced growth rate affects film morphology.

*Table S1: Physical vapor deposition conditions for plasma characterization and film growth*

| Film material | time (s)/ (ML) | substrate type | substrate temp (°C) | power type | power (W/kHz/μs) | Ar flow (sccm) | pressure (mTorr) |
|---|---|---|---|---|---|---|---|
| Mo | 4s; 5 ML | $Al_2O_3$; | RT | PDC | 90/65/0.4 | 25 | 10 |
| W | 4s; 5 ML | $Al_2O_3$; | RT | PDC | 90/65/0.4 | 25 | 10 |
| Mo | 16s; 5 ML | $Al_2O_3$; | RT | HIPIMS | 0.5kW/0.120/45 | 25 | 10 |
| W | 16s; 5 ML | $Al_2O_3$; | RT | HIPIMS | 0.5kW/0.120/45 | 25 | 10 |
| Mo | 2s; 5 ML | $Al_2O_3$; | RT | DC | 90/-/- | 25 | 10 |
| W | 2s; 5 ML | $Al_2O_3$; | RT | DC | 90/-/- | 25 | 10 |

The metal films deposited through DC, HIPMS, and pulsed DC sputtering were evaluated for optimized thickness, roughness, and conductivity. A Bruker AFM was used in tapping mode for step scans of all metal films to determine thickness values. Typical steps for tungsten films grown at different time under otherwise identical conditions are in Figure S2. Additionally, surface roughness is shown in Figure S3a for DC, HIPMS, and pulsed DC sputtering both at room temperature and while the substrate was heated to 500°C. An increased surface roughness was observed for films grown at higher temperature due to presumed crystallinity within the metal film. Four-point conductivity results for the same films are shown in Figure S3b. These results suggest that pulsed DC deposition at room temperature result in the smooth, thinnest films with the highest conductivity at sub-nm thicknesses.

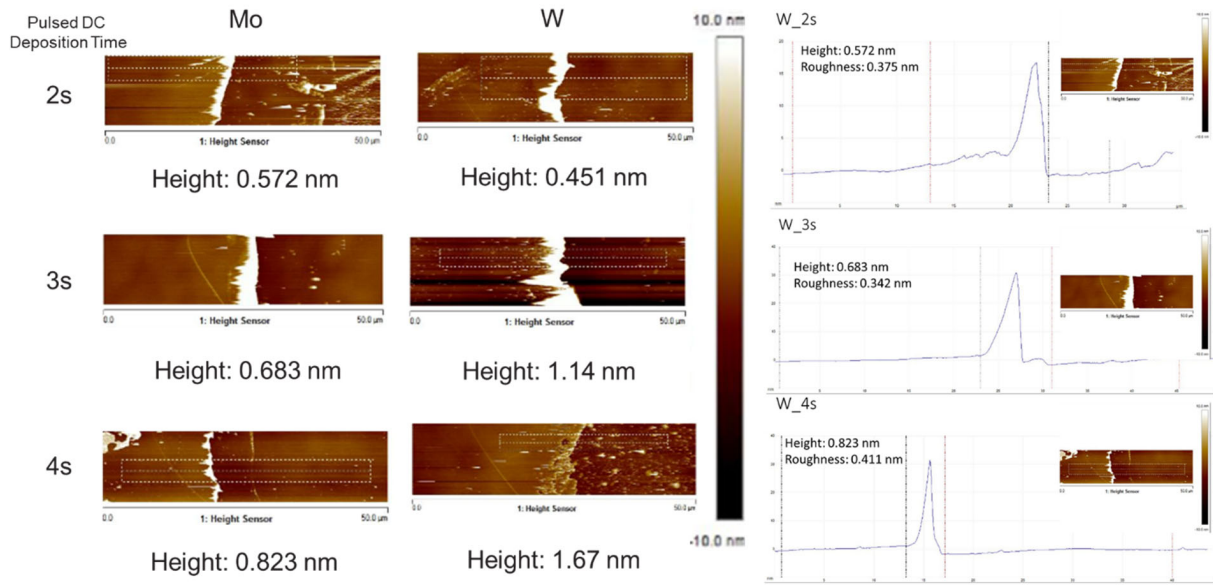

*Figure S2: Example AFM step height data and surface scans showing Mo and W film thickness and surface morphology for a range of deposition times of 2-4 s using Pulsed DC sputtering.*

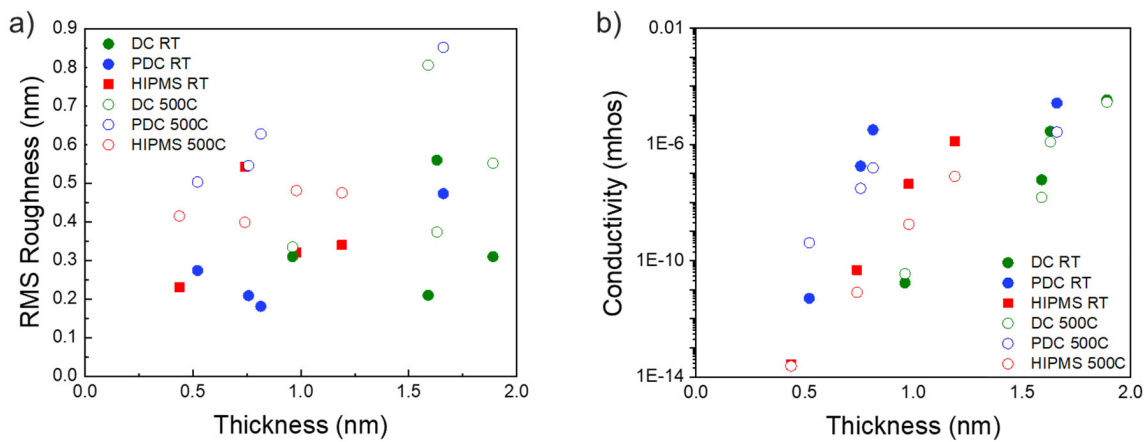

*Figure S3: a) Metal film roughness as a function of thickness and deposition type, and b) four-point conductivity of metal films as a function of thickness and deposition type.*

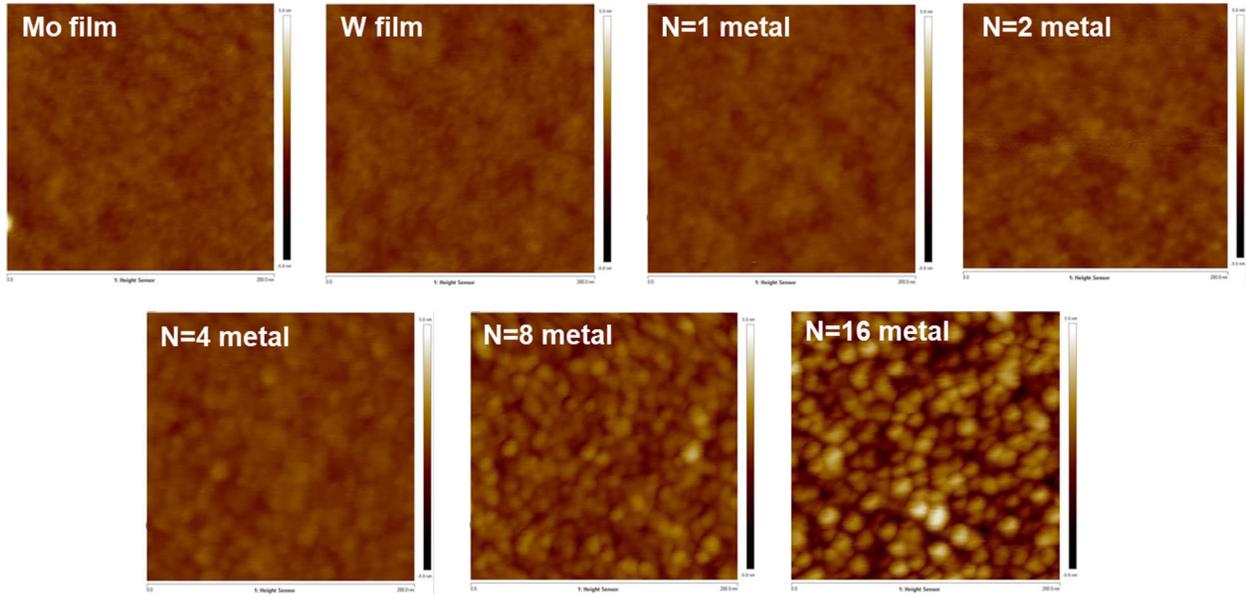

*Figure S4: AFM topography scans of preannealed metal films and heterostructures with scan dimensions being 2 μm x 2 μm. The surface roughnesses of each scan are the following: 0.256 nm, 0.246 nm, 0.228 nm, 0.278 nm, 0.257 nm, 0.528 nm, 1.10 nm.*

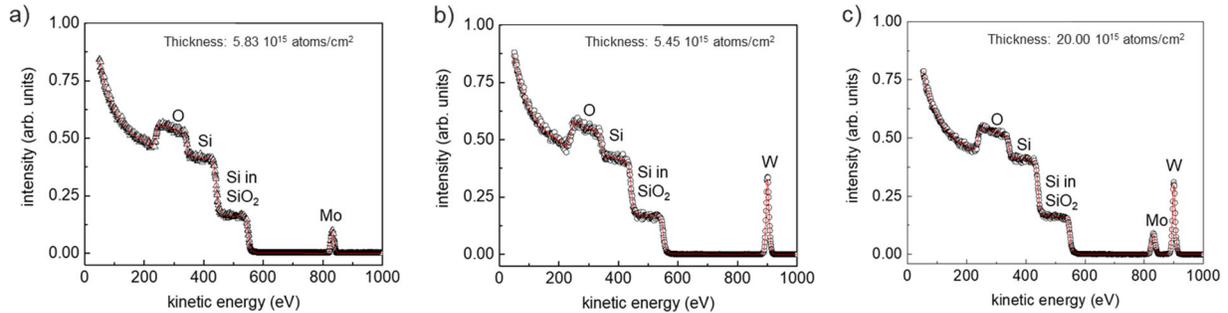

*Figure S5: Rutherford backscattering data and fitting parameters used for a) Mo films, b) W films, and Mo-W layered structure with N = 2. Red lines indicate fits of data using SIMNRA fitting and analysis software for the parameters shown.*

A Rutherford backscattering system employing a 4.6 MeV Tandetron accelerator for $^4$He ions with an incident energy of 2024 keV were used to generate the spectra shown in Figure S5 for monolithic Mo and W films, indicating approximately 5 monolayers of metal and also for a metal stack with N = 2 (Mo/W/Mo/W), indicated 20 layers of metal.

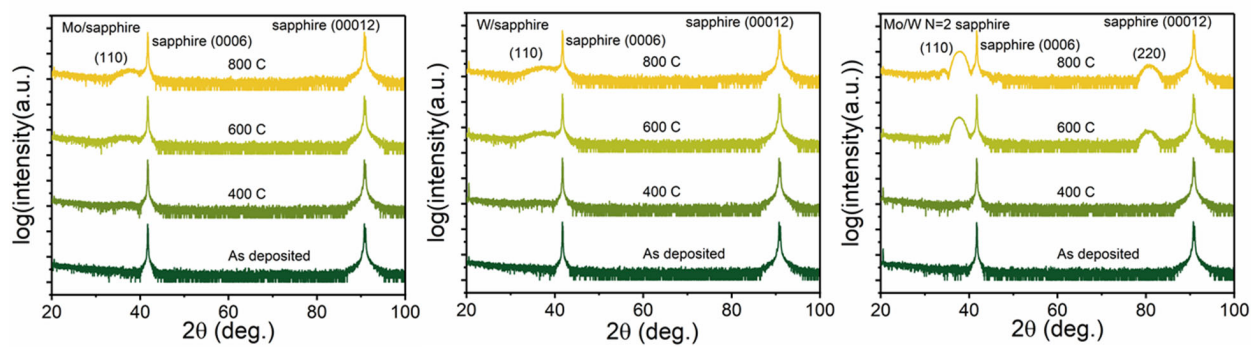

*Figure S6: X-ray diffraction of metallic Mo, W, Mo/W N=2 films on sapphire at various annealing temperatures without the presence of $H_2Se$ to evaluate crystal structure of the individual metal films. Metal films annealed at 500 Torr under a flow of $N_2$:$H_2$ (95%:5%).*

## Section S2: Annealing and conversion of metal films

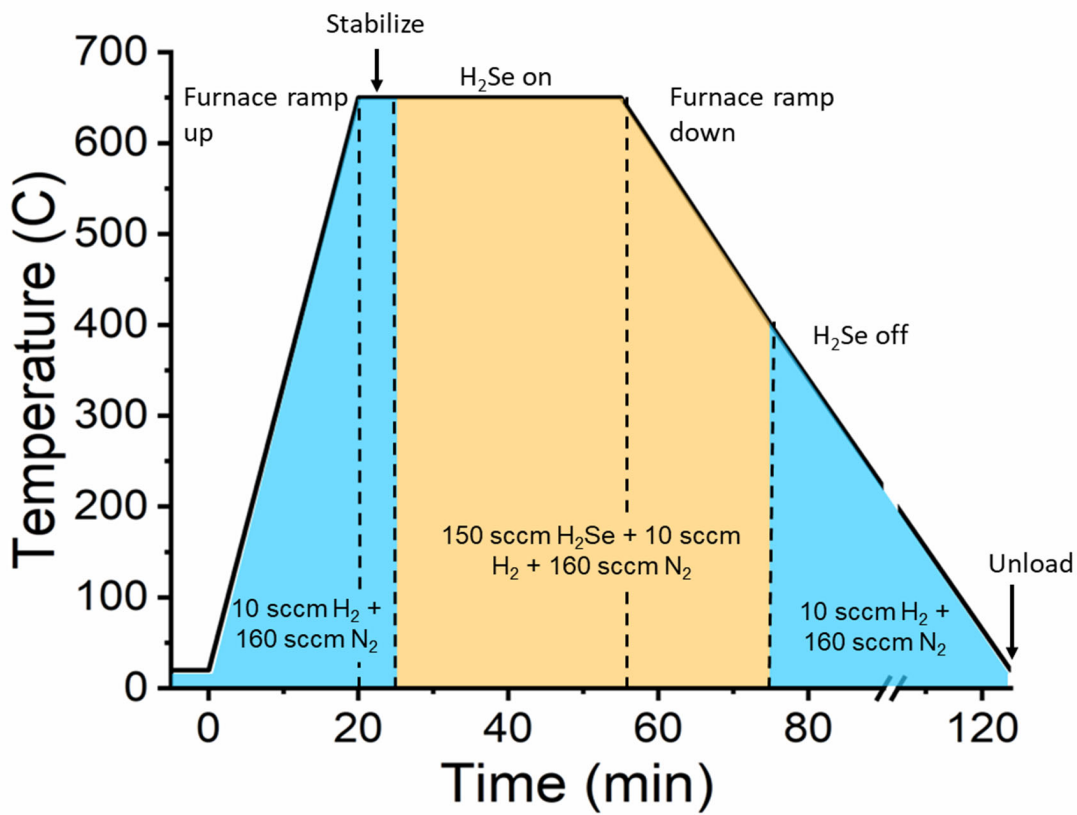

*Figure S7: Process run flow for $H_2Se$ annealing procedure.*

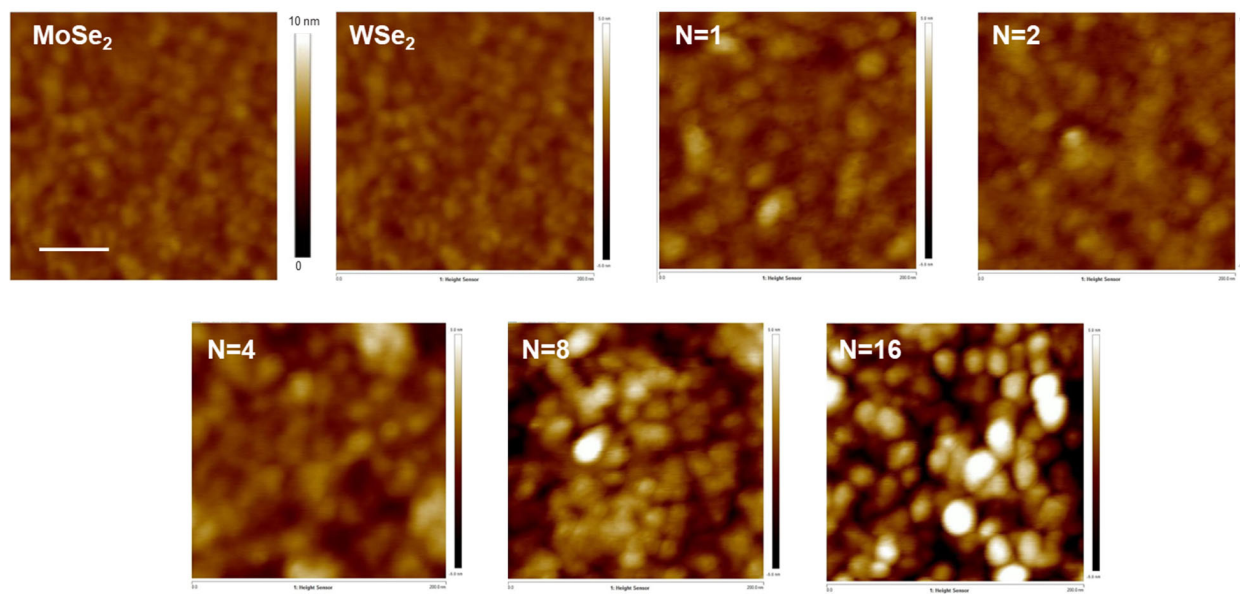

*Figure S8: AFM topography scans of TMD films and heterostructures after annealing in H$_2$Se with scan dimensions being 2 μm x 2 μm. The surface roughnesses of each scan are the following: 0.631 nm, 0.375 nm, 0.631 nm, 0.549 nm, 0.967 nm, 1.51 nm, 2.28 nm.*

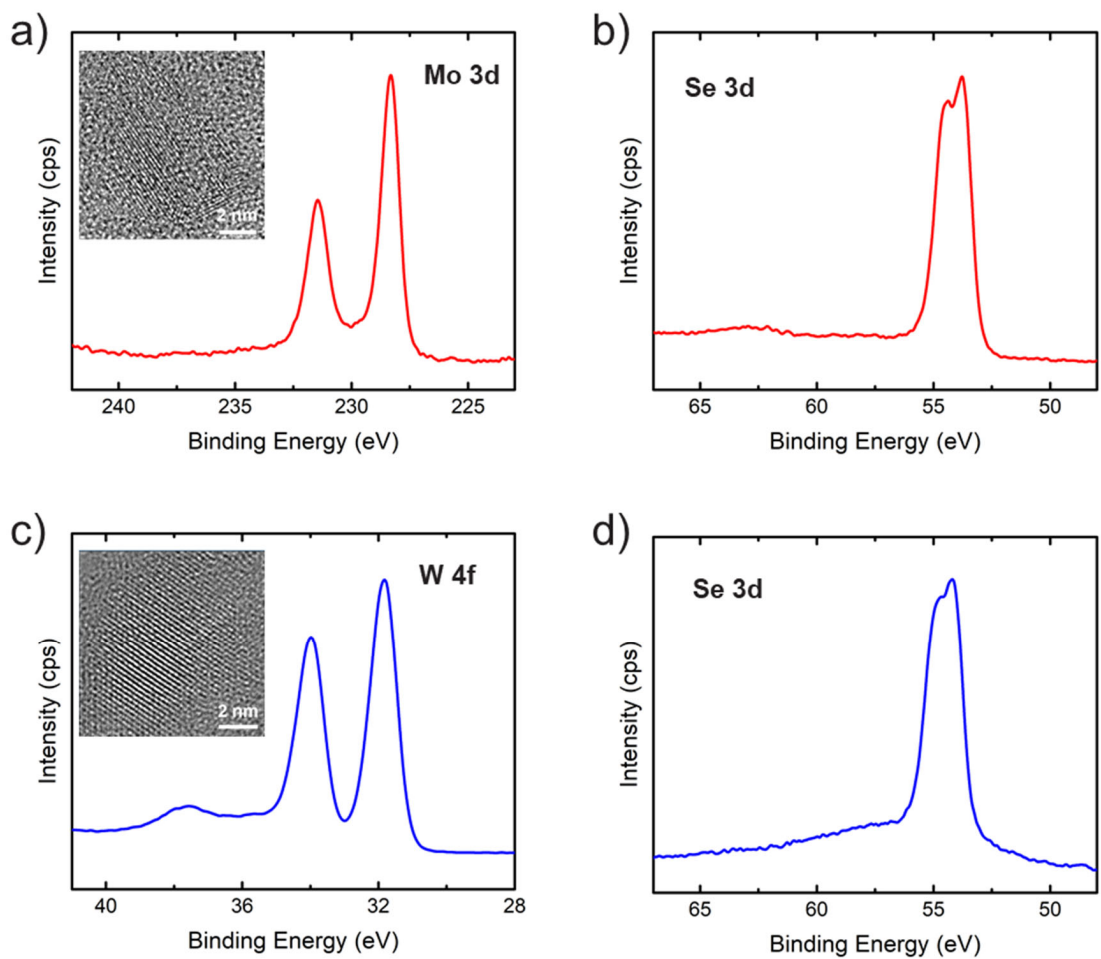

*Figure S9: XPS spectra of a) few-layer MoSe$_2$ Mo3d with inset TEM image of few-layer crystalline film and b) Se 3d, c) few layer WSe$_2$ W 4f with inset TEM image of few-layer crystalline film and d) Se 3d.*

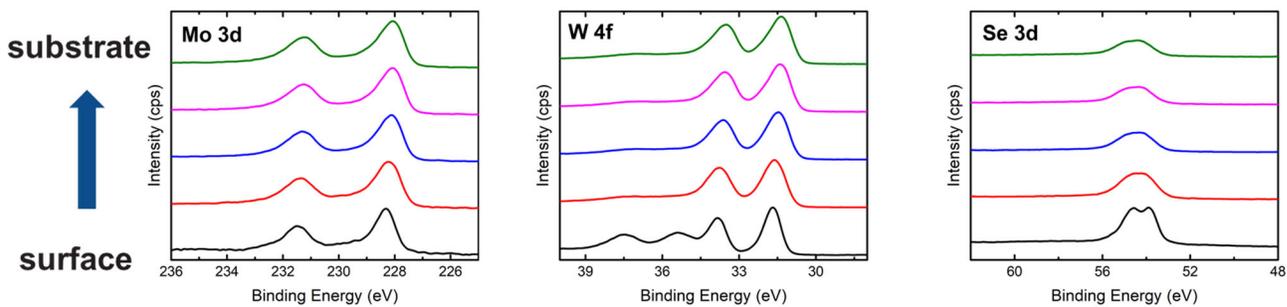

*Figure S10: Depth-profile XPS scans of Mo3d, W4f, and Se3d of N=8 superlattice structure, demonstrating full selenization throughout the structure and some evidence of surface oxidation (black curve).*

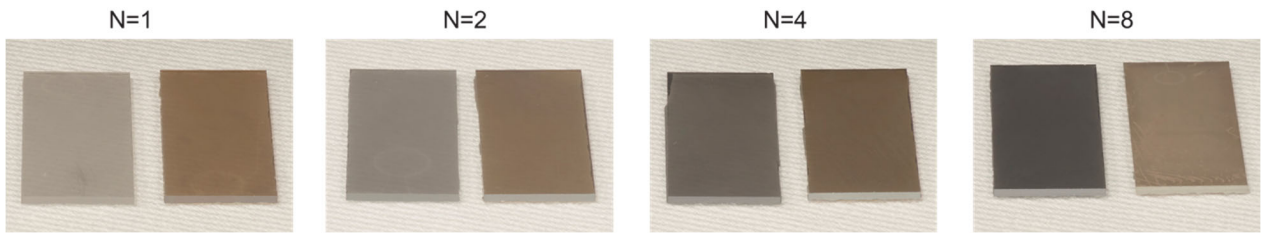

*Figure S11: Close-up optical images of N=1, 2, 4, and 8 superlattice samples on 2 cm x 1 cm sapphire wafers*

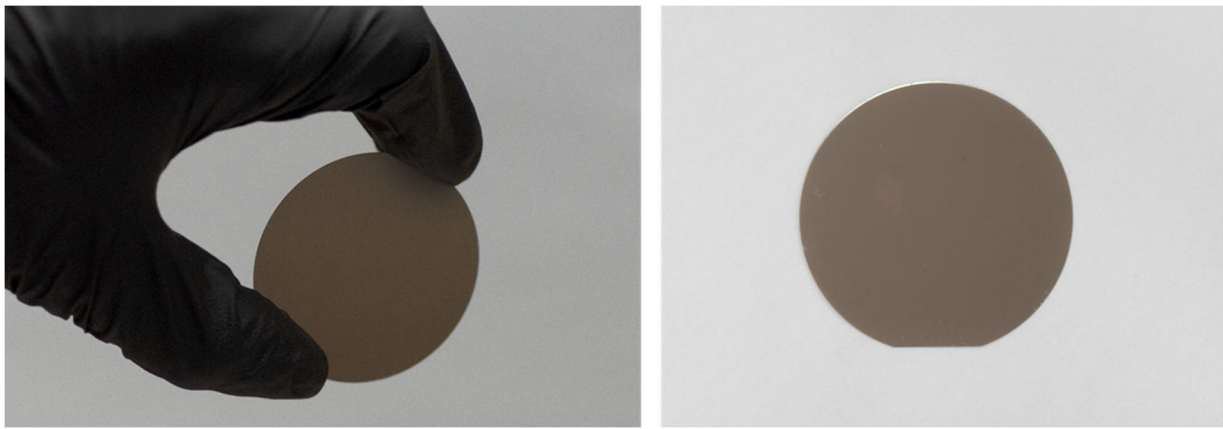

*Figure S12: Optical image of 2" wafer of N=2 superlattice film on sapphire*

**Section S3: Raman and polarized Raman of superlattice films as a function of temperature and substrate**

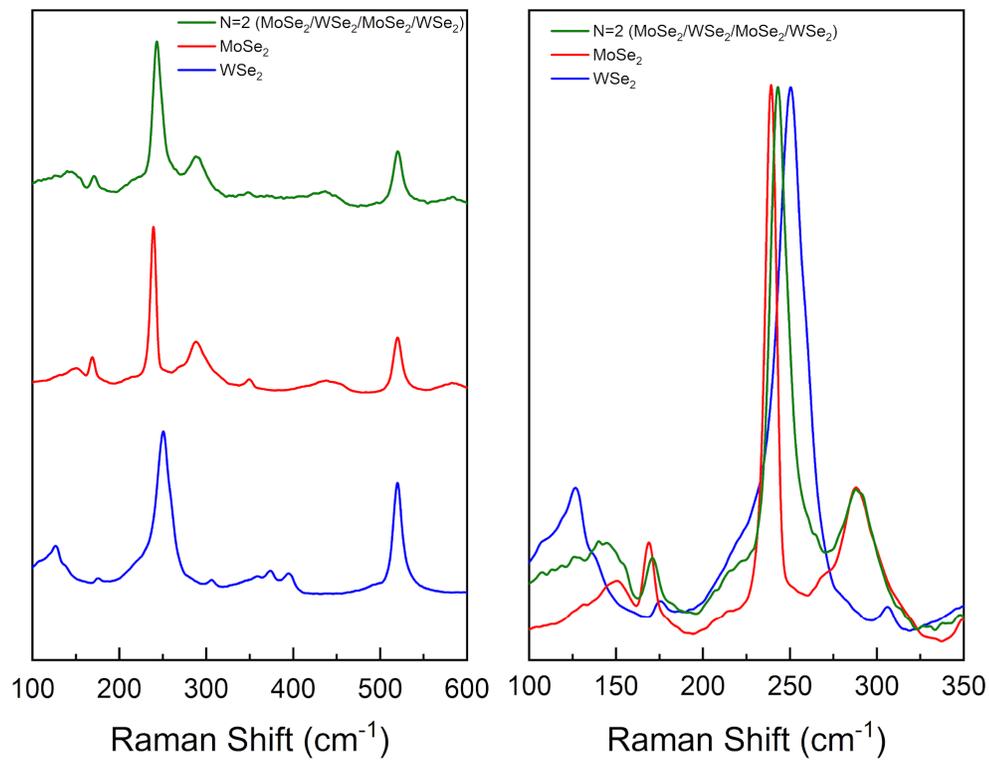

*Figure S13: High resolution Raman scans of MoSe$_2$, WSe$_2$, and N=2 superlattice structure on sapphire substrates.*

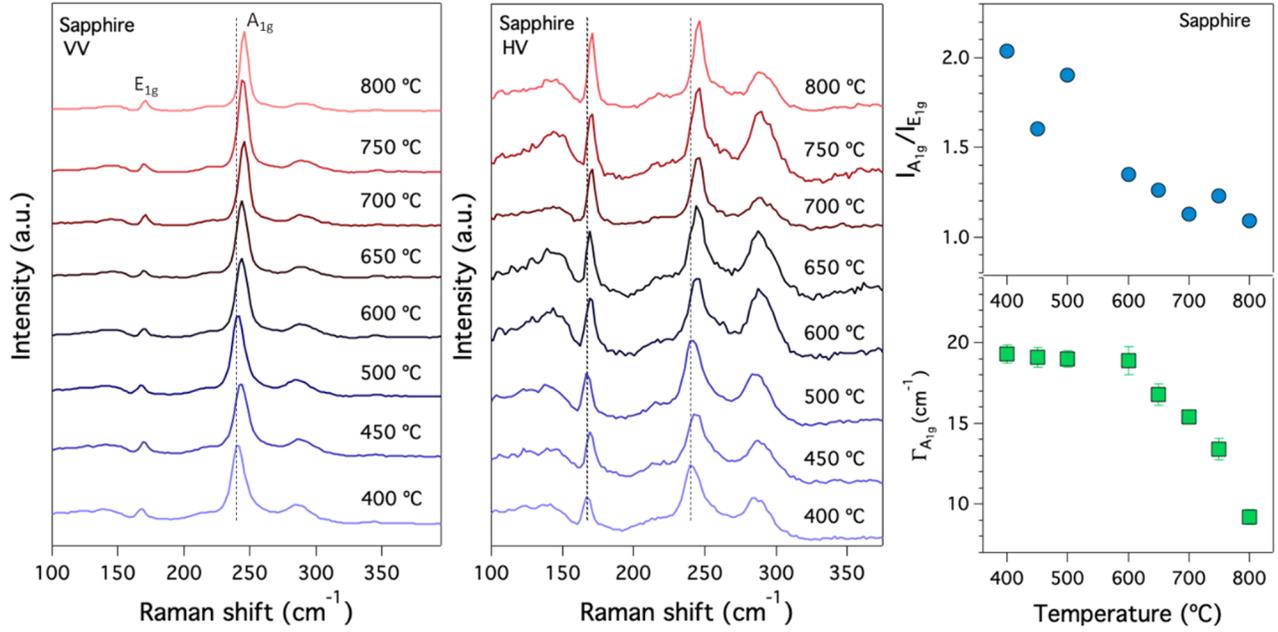

*Figure S14: Polarized Raman on N=2 superlattice films annealed in H₂Se at different temperatures on sapphire substrates including VV and HV polarization as well as the intensity ratio of the A$_{1g}$ and E$_{2g}$ peaks and FWHM of A$_{1g}$ peak as a function of annealing temperature.*

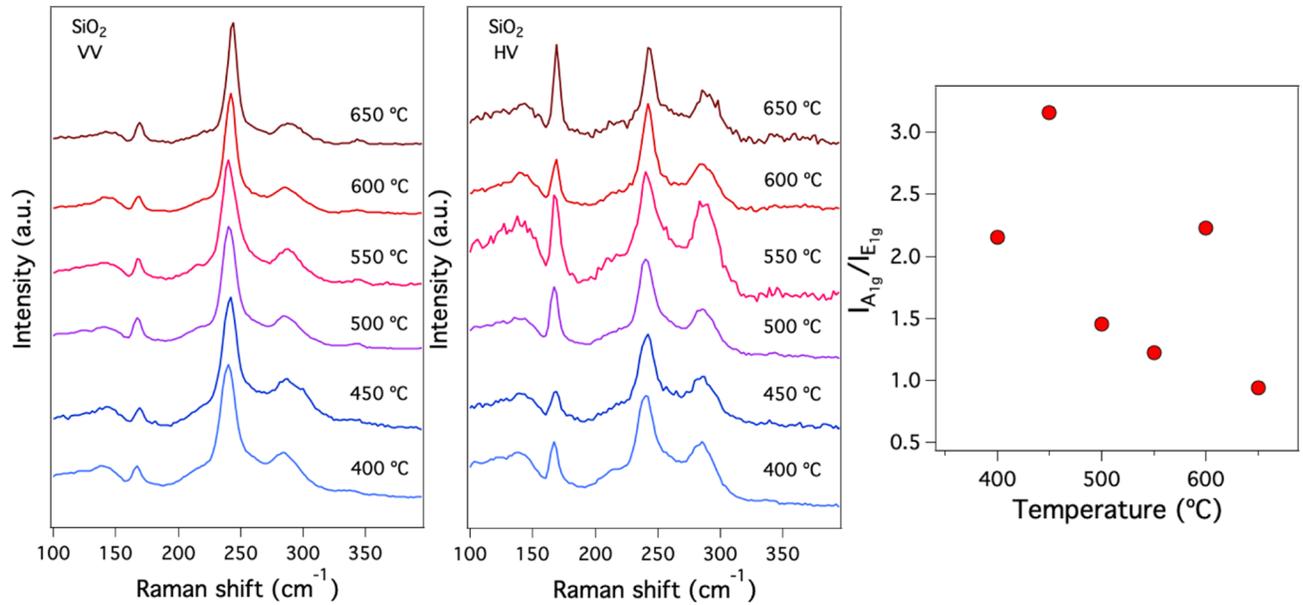

*Figure S15: Polarized Raman on N=2 superlattice films annealed in H₂Se at different temperatures on Si/SiO₂ substrates including VV and HV polarization as well as the intensity ratio of the A$_{1g}$ and E$_{2g}$ peaks as a function of annealing temperature.*

**Section S4: Structural characterization of annealed films via XRD and SHG**

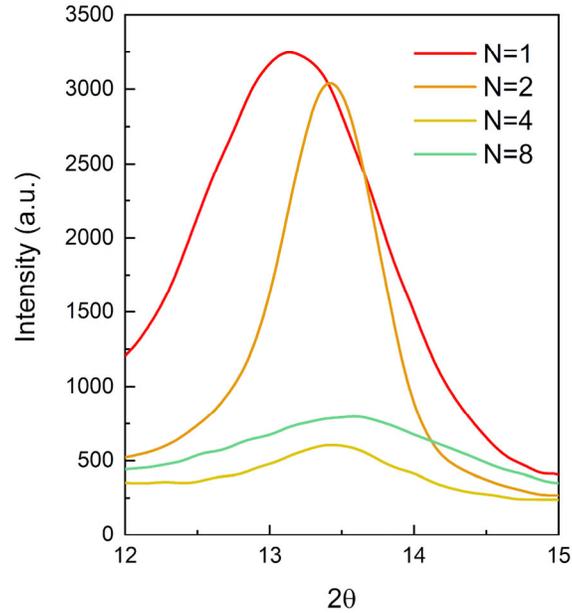

*Figure S16: XRD spectra for N=1, 2, 4, and 8 superlattices annealed at 600°C in $H_2Se$.*

To further understand the structural properties of the films, we apply scanning second harmonic generation to estimate the grain size of few-layer $MoSe_2$, $WSe_2$, N=1 heterostructure as well as N=2 heterostructures annealed at 400 °C and 800 °C, as shown in Figure S17 and S18. A pulsed laser centered at 800 nm with an average power of 4 mW (MaiTai SP, 84 MHz) is focused to a 3-μm-diameter spot, providing an optical resolution of d=2 μm at the 400 nm imaging wavelength. The images exhibit nearly uniform SHG intensity that are insensitive to laser polarization, characteristic of a polycrystalline film with domain size much smaller than the optical resolution. Following numerical simulation, the spatial variation of SHG intensity of randomly aligned and stacked domains is expected to be $0.17(s/d)^2$ when normalize by the maximum SHG intensity of a single crystalline monolayer $WSe_2$, where s is the average domain size. We observed nearly an-order-of-magnitude increase in SHG spatial variance, indicating the growth of average lateral domain size from 20 nm to 60 nm after 800 °C annealing, consistent with STEM and XRD results.

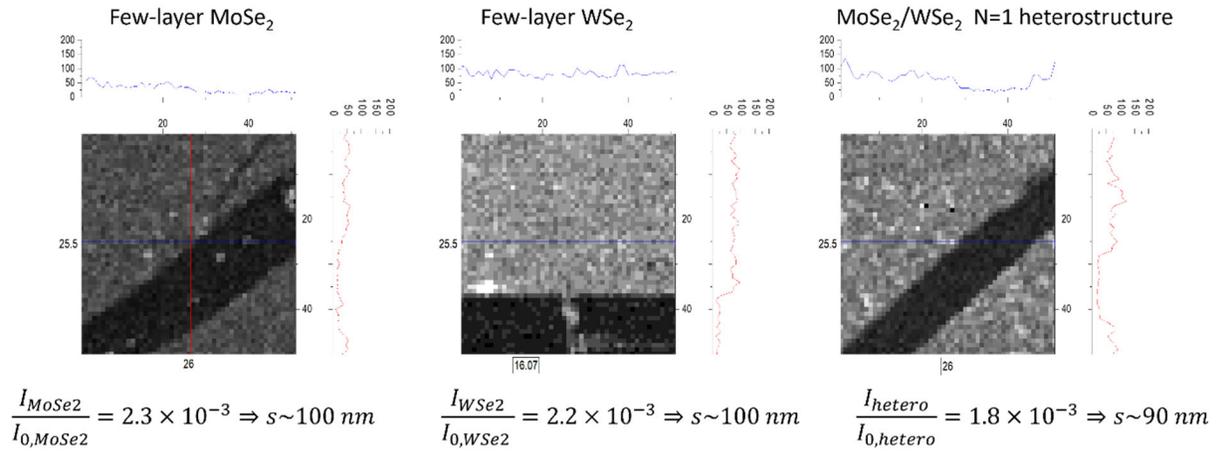

*Figure S17: SHG grain size measurements of few-layer MoSe$_2$ and WSe$_2$ and N=1 heterostructure.*

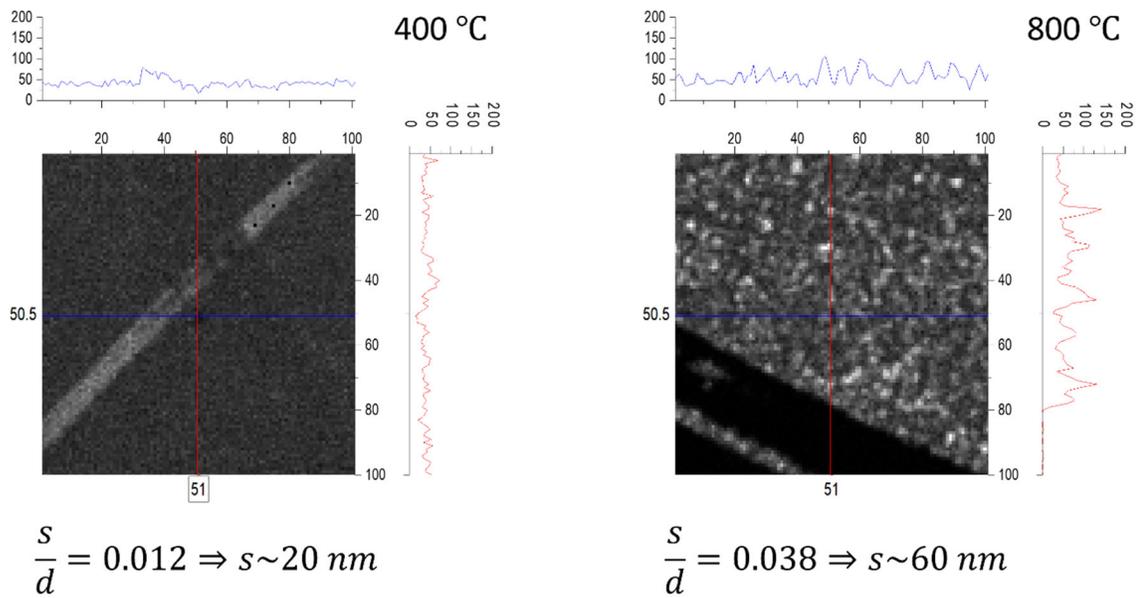

*Figure S18: SHG of N=2 heterostructures.*

## Section S5: TEM and EELS characterization of annealed films

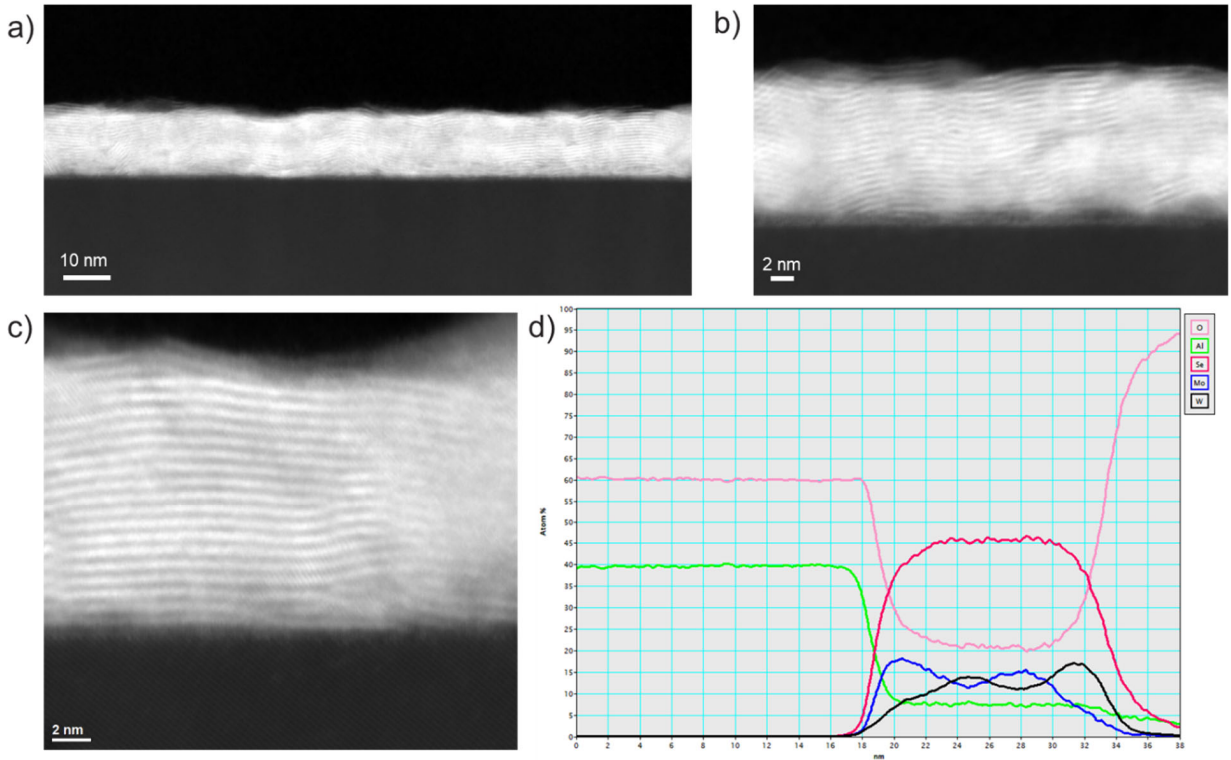

*Figure S19: N=2 600°C Superlattice STEM and EELS data. a), b), and c) various TEM images of superlattice structure with d) EELS cross-section data from image c showing segregation of Mo and W.*

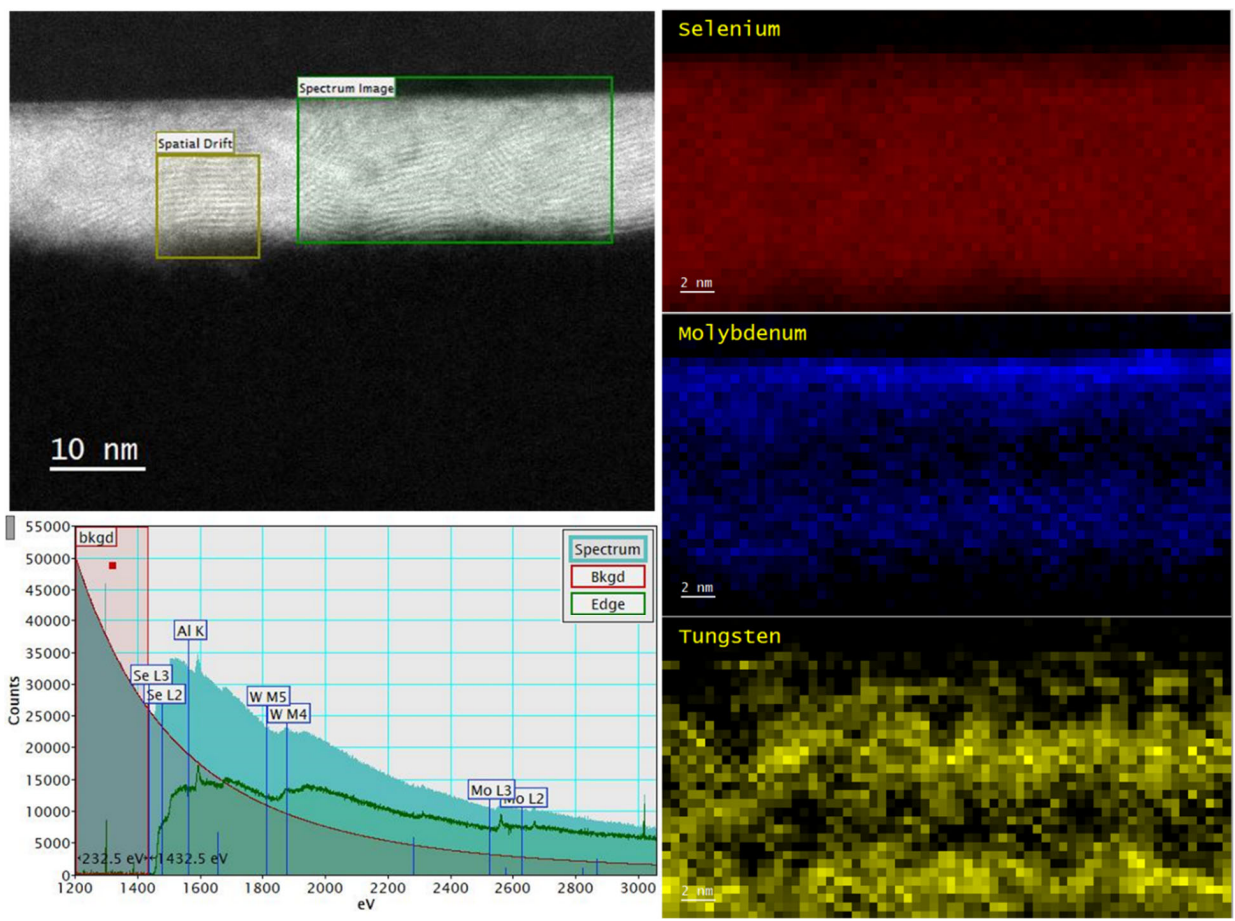

*Figure S20: Cross-sectional STEM view of N=2 400C annealed mixed orientation superlattice structure with corresponding EELS spectrum and mapping.*

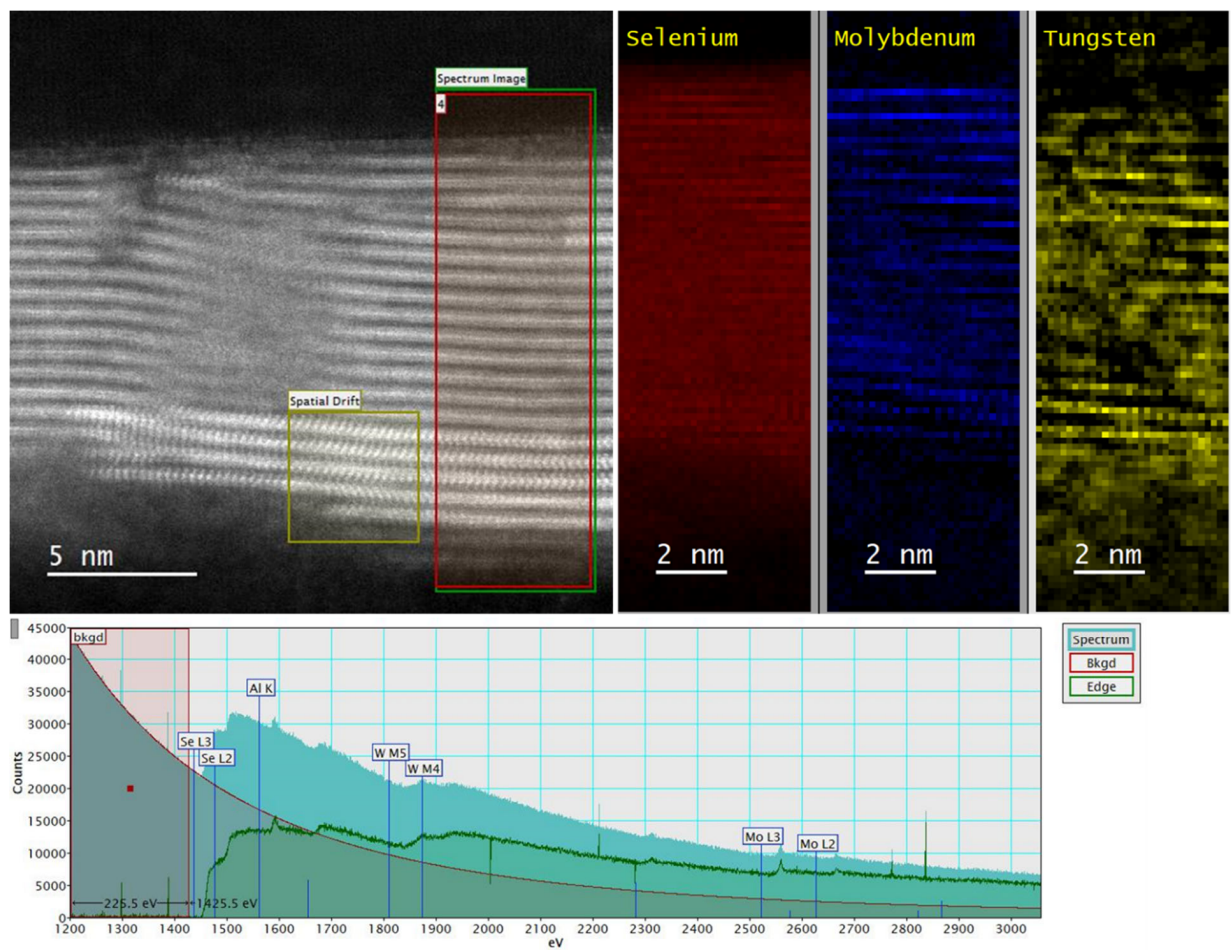

*Figure S21: Cross-sectional STEM view of N=2 800°C annealed horizontal orientation alloyed structure with corresponding EELS spectrum and mapping.*

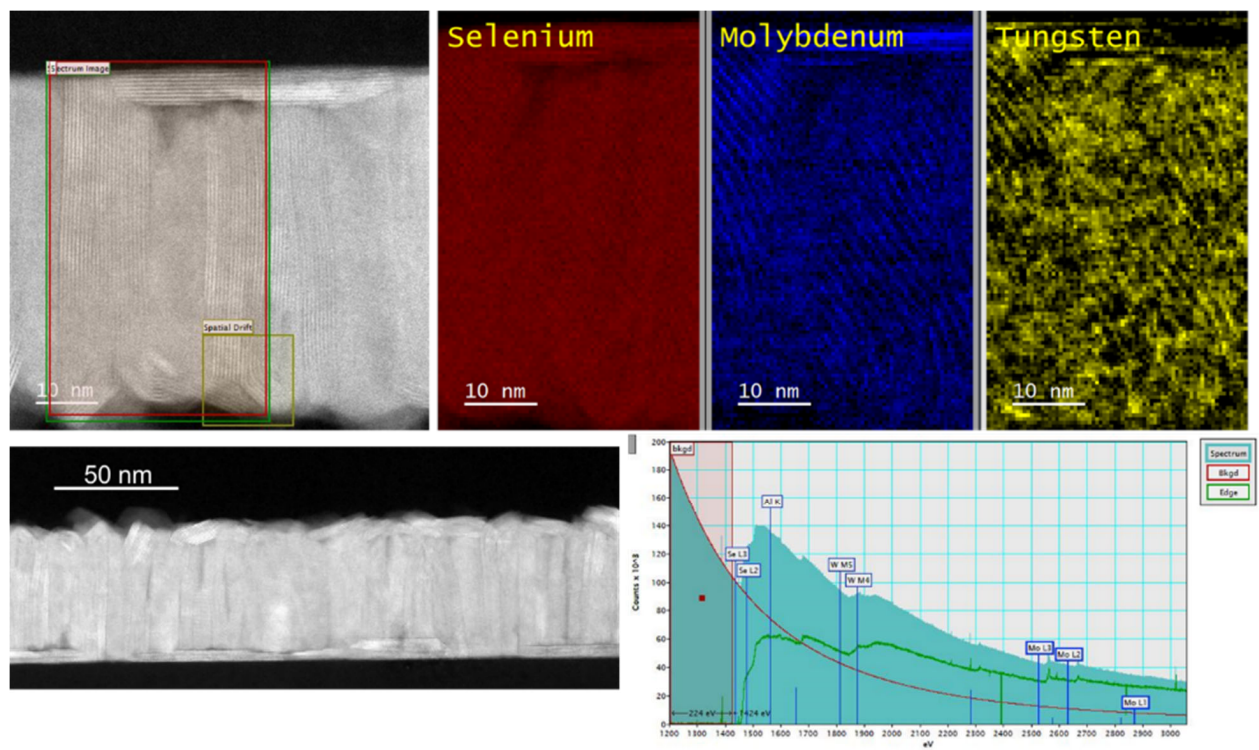

*Figure S22: STEM and EELS analysis of vertically oriented N=8 alloyed structure annealed in $H_2Se$ at 800 °C.*

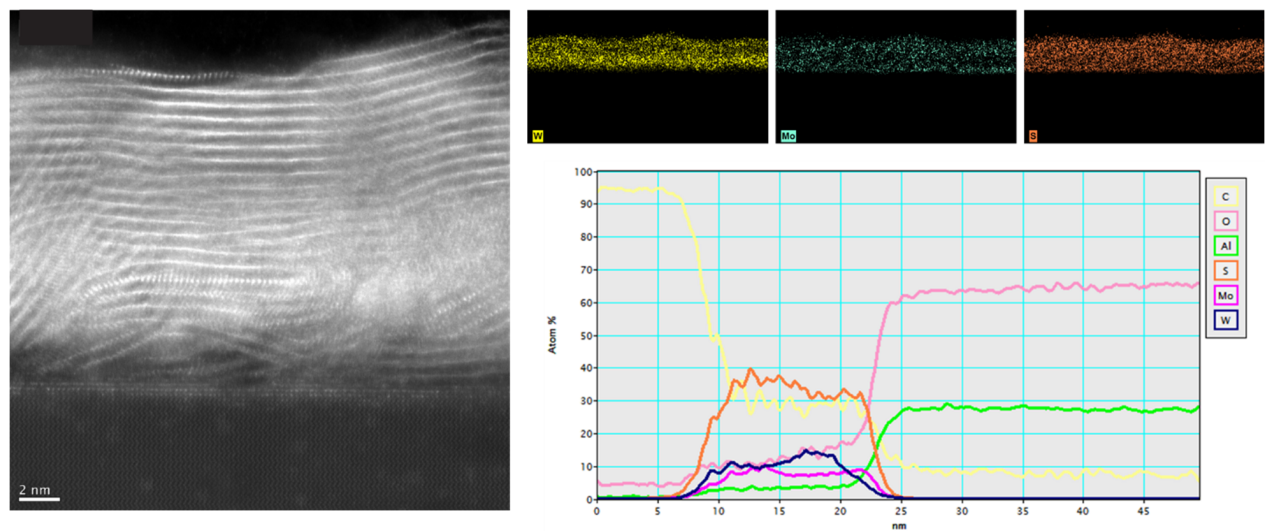

*Figure S23: STEM and EELS analysis of $MoS_2$/$WS_2$ superlattice films annealed in $H_2S$ at 650 °C, showing some intermetallic separation with potentially increased mixing in the sulfide structures.*

## Section S6: Structural, optical and electronic simulations of superlattices

The lattice parameters and space group of $WSe_2$, $MoSe_2$, monolayer $WSe_2$/monolayer $MoSe_2$ superlattice ($1WSe_2/1MoSe_2$), and five-layer $WSe_2$/five-layer $MoSe_2$ superlattice ($5WSe_2/5MoSe_2$) structures obtained with DFT are presented in Table S2, together with experimental data.[1] Comparing the theoretical data with the experimental one for $WSe_2$ and $MoSe_2$, we found a slight percentage deviation of 0.43% and 0.21% for the lattice parameter a, and 3.10% and 3.49% for the lattice parameter c, respectively. The lattice mismatch for parameter a between $WSe_2$ and $MoSe_2$ is less than 1%. Comparing $1WSe_2/1MoSe_2$ and $5WSe_2/5MoSe_2$ superlattices to $WSe_2$ or $MoSe_2$, the parameter a decrease ~0.5% and ~0.4%, respectively. While the parameter c of $1WSe_2/1MoSe_2$ is ~1.6% lower than that of $WSe_2$ or $MoSe_2$, the parameter c of $5WSe_2/5MoSe_2$ is ~3.5% lower than that of 5-layer $WSe_2$ or 5-layer $MoSe_2$. These results suggests that the interaction between the $WSe_2$ layer and $MoSe_2$ layer in the superlattice causes a contraction in both parallel and perpendicular directions of the layers.

Table S2: Structural and electronic properties of $WSe_2$, $MoSe_2$, $1WSe_2/1MoSe_2$, and $5WSe_2/5MoSe_2$ structures after the DFT/PBE/PAW geometry optimization.

|  | DFT/PBE/PAW | | | Experimental data | | | Space group |
| --- | --- | --- | --- | --- | --- | --- | --- |
|  | a (Å) | c (Å) | Bandgap (eV) | a (Å) | c (Å) | Bandgap (eV) |  |
| $WSe_2$ | 3.294 | 13.352 | 1.070 | 3.280 | 12.950 | 1.210 | $P6_3/mmc$ |
| $MoSe_2$ | 3.295 | 13.351 | 0.989 | 3.288 | 12.900 | 1.090 | $P6_3/mmc$ |
| $1WSe_2/1MoSe_2$ | 3.279 | 13.139 | 0.936 | --- | --- | --- | $P\underline{6}m2$ |
| $5WSe_2/5MoSe_2$ | 3.283 | 64.413 | 0.722 | --- | --- | --- | $P\underline{6}m2$ |

According to the geometrical parameters presented in Table S2, an interlayer spacing (transition metal layer spacing) of ~6.67 Å can be calculated for $WSe_2$ and $MoSe_2$, around 3% larger than the experimental results.[2–4] For $1WSe_2/1MoSe_2$, the interlayer spacing is 6.46 Å, lower than the observed for $WSe_2$ and $MoSe_2$. For the $5WSe_2/5MoSe_2$, we observed an interlayer spacing of 6.43 Å, 6.44 Å, and 6.44 Å between the $WSe_2$ and $MoSe_2$, $WSe_2$ and $WSe_2$, and $MoSe_2$ and $MoSe_2$, respectively.

The electronic band structures of $WSe_2$, $MoSe_2$, and $1WSe_2/1MoSe_2$ are presented in Figure S24. All of them have an indirect bandgap that occurs from the Γ point to a k point approximately at the middle of the K-Γ path. According to the values of bandgap listed in Table S2, they can be sorted in descending order as: $WSe_2$, $MoSe_2$, $1WSe_2/1MoSe_2$, $5WSe_2/5MoSe_2$. The deviations between the bandgaps of $WSe_2$ and $MoSe_2$ obtained by PBE and the experimental data are found to be -9.26% and -11.57%,[1] respectively, which are within the range of the average error for bandgaps estimation with DFT (~10%).[5,6] In general, our PBE results show an average error of -10.41% in comparison with the available experimental bandgaps. If we correct the theoretical bandgaps of 1WSe$_2$/1MoSe$_2$ and 5WSe$_2$/5MoSe$_2$ heterostructures based on this average error, we could expect to obtain their experimental bandgaps of around 1.033 eV and 0.797 eV, respectively.

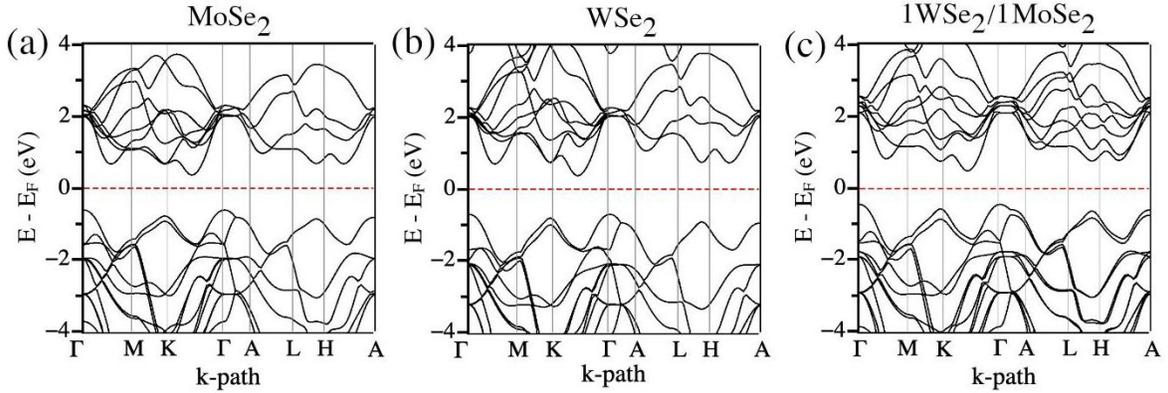

Figure S24. Theoretical electronic band structure of the (a) WSe$_2$, (b) MoSe$_2$, (c) 1WSe$_2$/1MoSe$_2$.

The PDOS and the square of the wave function of VBM and CBM of 1WSe$_2$/1MoSe$_2$ are displayed in Figure S25. They look similar to these of 5WSe$_2$/5MoSe$_2$, except that the VBM spreads along all the superlattice and mostly in WSe$_2$ layer, the CBM is localized only in the MoSe$_2$ layer.

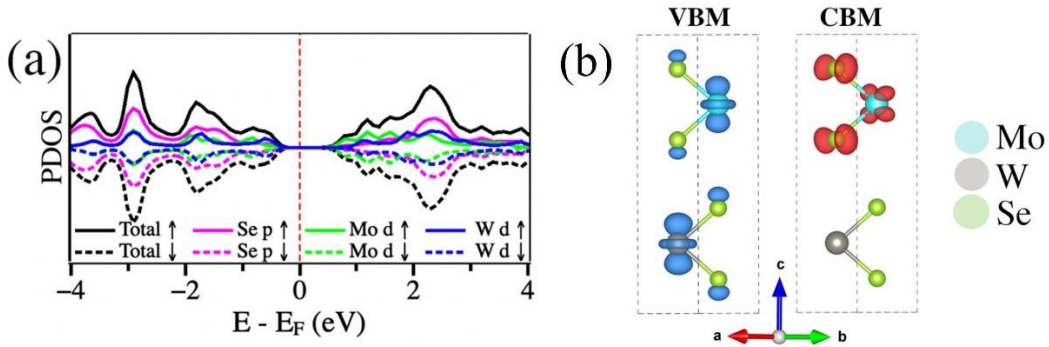

Figure S25. (a) PDOS and (b) the square of the wave function of VBM and CBM of 1WSe$_2$/1MoSe$_2$.

Figure S26a shows the $E_{CB} - E_{VB}$ curves for the superlattice film and the possible transitions related to CPs A to D. These possible electronic transitions in the Brillouin zone are located at the tangent points ($\nabla(E_{CB} - E_{VB})=0$) of the curves according to the Joint Density of States (JDOS) (17). The energy of the CP A (1.39 eV) is close to the energy of a VB$_3$ to CB$_2$ transition inside the M → K path (see Figure S26b). However, three more tangent points that correspond to a $VB_3$ to $CB_1$, $VB_3$ to $CB_3$, and $VB_4$ to $CB_3$ transitions inside the M → K path also have energy very close to that of CP A. This suggests that due to the very close energy among these four tangent points (differences in the order of 0.001 eV), the CP A could be related to any of them or a combination of them. As for the CP B (1.74 eV), this energy corresponds to a $VB_1$ to $CB_1$ electronic transition at a k-point inside the K → Γ path (see Figure S26c). There are four more tangent points above the corresponding to the CP B, with energy differences of 0.01 eV among each other. In Figure S26d, it is possible to see that the CP C occurs at a $VB_3$ to $CB_1$ electronic transition in a k-point inside the Γ → M path. There are at least 8 more tangent points close to the one assigned as the C. As for CP D (2.64 eV), it can be assigned as a $VB_2$ to $CB_1$ electronic transition in a k-point inside the Γ → M path (see Figure S26e). It is also possible to notice other tangent points around the CP D.

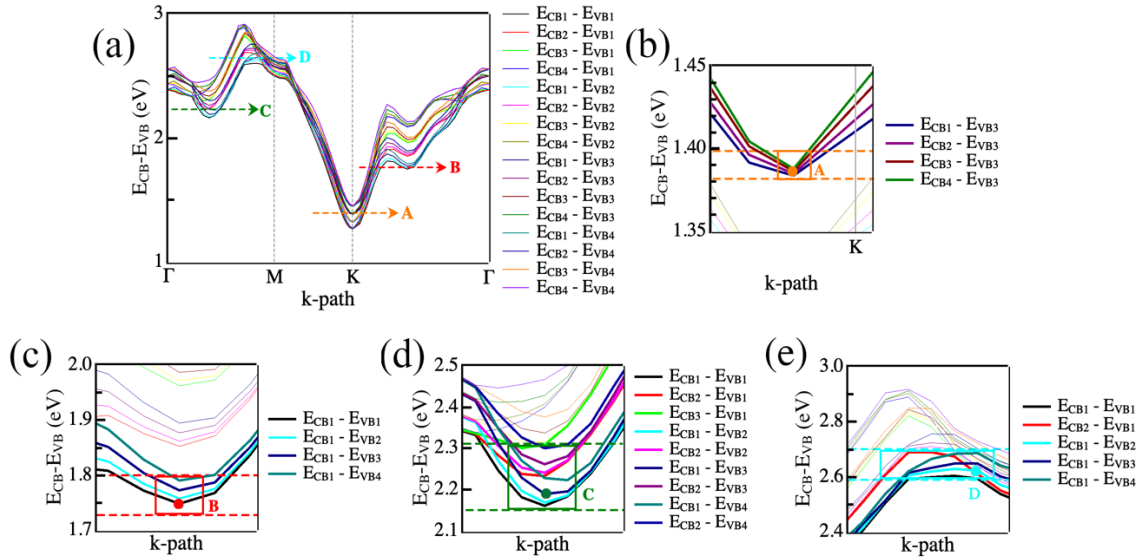

*Figure S26: (a) $E_{CB} - E_{VB}$ curves for the energy differences for the last four and first four VBs and CBs with the k-path in the Brillouin zone for superlattice film. Zoomed view of the position of the CPs (b) A, (c) B, (d) D, and (e) D in the $E_{CB} - E_{VB}$ curves.*

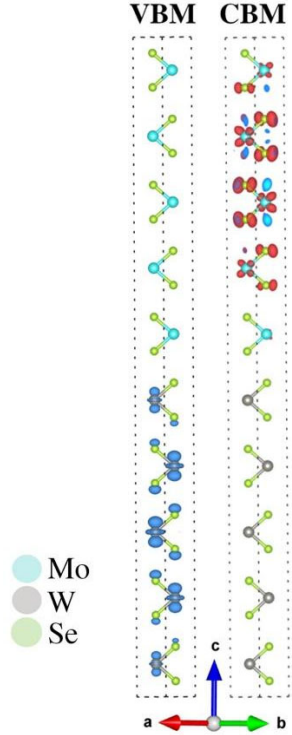

*Figure S27: square of the wave function of VBM and CBM for 5L MoSe2/5L WSe2 superlattices.*

## Section S7: Optical characterization of superlattice and alloyed films

The complex refractive index (Figure 1f,h and Figure S28b and c and complex dielectric function (Figure 4a and S29a) for individually grown MoSe2 and WSe2 layers and MoSe2/WSe2 superlattices (N = 1, 2, 4, 8) at their optimal growth temperatures are shown in Figure S29. There is a clear increase in the $\varepsilon_1$ (Figure S28a) and $\varepsilon_2$ (Figure 4a) values for energies $E \leq \sim 2.5$ eV for the N = 1, 2 superlattices compared to the individually grown MoSe2 and WSe2 layers. We note that these responses suggest distinct superlattice structures as opposed to the alloys that form for N = 4, 8. This increase is not explainable by the effective medium approximation, valid for thickness $t \ll \lambda$, which limits the relative permittivity of the system to a volume-fraction weighted average of the individual components.[7] As such, the change in the optical constants is likely due to a commensurate change in the electronic structure of the superlattice.

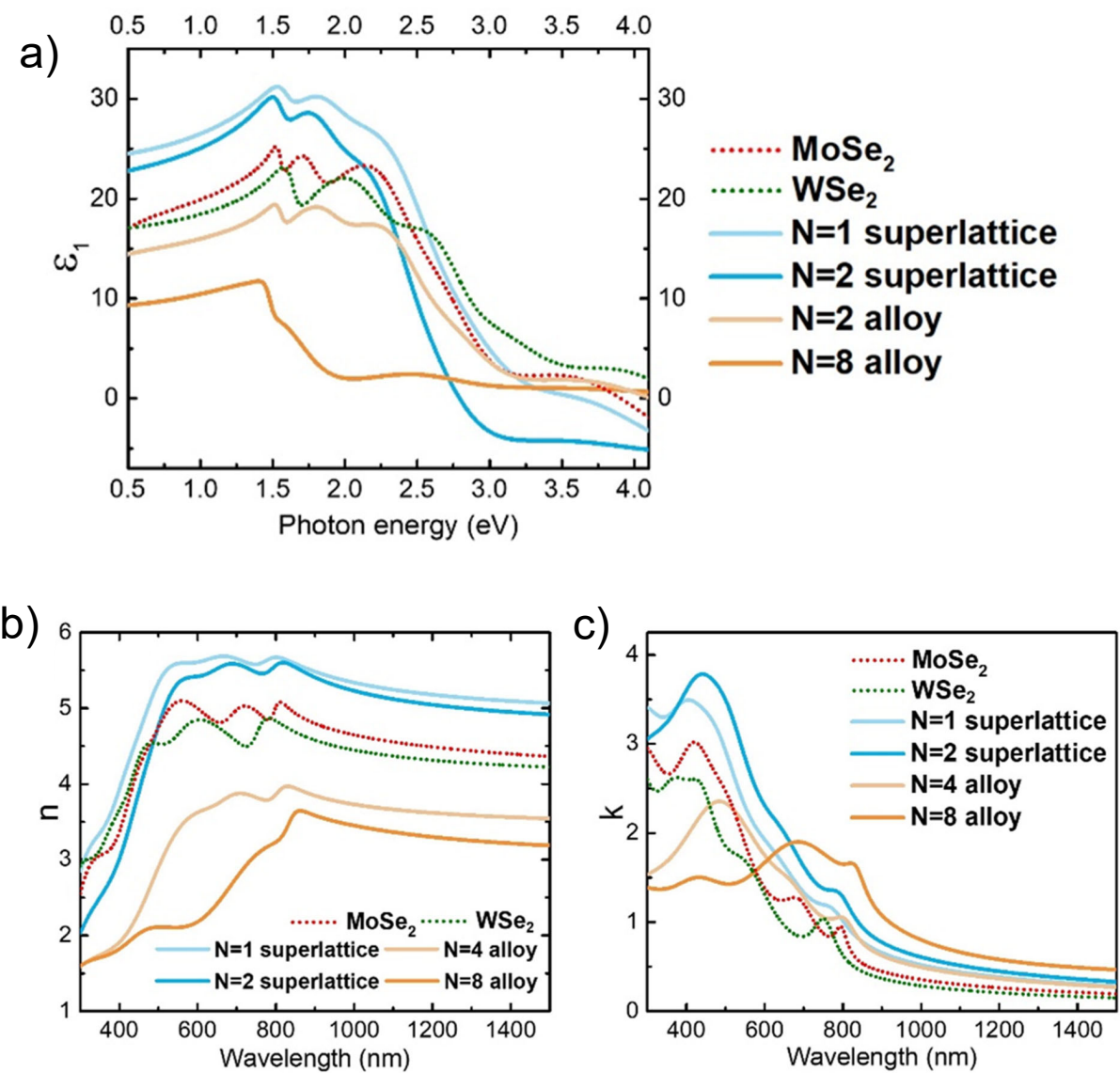

*Figure S28: a) ε1, b) n and c) k values of MoSe$_2$ and WSe$_2$ individual few-layer films, N=1 and N=2 superlattice, N=2 alloy and N=8 vertical alloy.*

*Table S3*: Lorentz multi-oscillator fit parameters for the MoSe$_2$, WSe$_2$, N=1 superlattice, N=2 superlattice, N=2 alloy, and N=8 alloy films used for ellipsometry optical dispersion data analysis to determine $n$, $k$, $\varepsilon_1$, $\varepsilon_2$, and $\alpha$ values.

| Lorentz Parameter[a] | Oscillator 1 | Oscillator 2 | Oscillator 3 | Oscillator 4 | Oscillator 5 | Oscillator 6 |
|---|---|---|---|---|---|---|
| *MoSe$_2$ (N=0) film (MSE # = 2.8 with a derived total film thickness of 3.4 nm)* | | | | | | |
| Amp[b] | 1.405821 | 1.251569 | 1.947515 | 8.322865 | 9.742136 | - |
| Br[c] | 0.1848 | 0.3883 | 0.3425 | 1.1465 | 4.2885 | - |
| En[d] | 1.685 | 2.075 | 2.303 | 2.874 | 4.815 | - |
| *WSe$_2$ (N=0) film (MSE = 2.2 with a derived total film thickness of 3.5 nm)* | | | | | | |
| Amp | 2.540689 | 7.864161 | 11.820073 | 3.095070 | 8.002196 | 8.586362 |
| Br | 0.2286 | 0.7492 | 0.7665 | 0.6057 | 3.1289 | 4.2455 |
| En | 1.686 | 2.287 | 2.834 | 3.301 | 4.297 | 4.982 |
| *N=1 superlattice film (MSE 3.2 = with a derived total film thickness of 7.0 nm)* | | | | | | |
| Amp | 2.782981 | 5.939684 | 17.876764 | 12.154141 | 11.480013 | 5.935669 |
| Br | 0.1970 | 0.6019 | 0.9428 | 1.1502 | 1.8801 | 3.4169 |
| En | 1.597 | 1.959 | 2.537 | 3.022 | 4.056 | 5.388 |
| *N=2 superlattice film (MSE 4.3 = with a derived total film thickness of 13.8 nm)* | | | | | | |
| Amp | 3.515517 | 7.152371 | 16.473323 | 12.454491 | 10.650405 | - |
| Br | 0.1677 | 0.5218 | 0.7907 | 1.0180 | 3.1713 | - |
| En | 1.561 | 1.900 | 2.394 | 2.767 | 3.989 | - |
| *N=2 alloy film (MSE 5.2 = with a derived total film thickness of 16.5 nm)* | | | | | | |
| Amp | 2.405567 | 4.001543 | 9.996008 | 8.599408 | 9.288371 | - |
| Br | 0.1106 | 0.4546 | 0.7194 | 0.9723 | 2.3156 | - |
| En | 1.562 | 1.941 | 2.488 | 2.954 | 4.260 | - |
| *N=8 alloy film (MSE 8.5 = with a derived total film thickness of 65.9 nm)* | | | | | | |
| Amp | 2.630484 | 8.372043 | 2.498634 | 3.233971 | 1.273188 | - |
| Br | 0.1122 | 0.7798 | 1.1972 | 3.8556 | 4.6322 | - |
| En | 1.481 | 1.723 | 2.779 | 4.549 | 8.358 | - |

[a]Lorentz parameters shown here are those used in the computational software and are related to Equation S# as Amp for amplitude where $f = Amp \cdot Br \cdot En$ … in units of eV$^2$, Br for oscillator peak broadness where $\gamma = Br$ in units of eV, and En for the energy of the oscillator peak in eV.

The measured normal incidence reflectance spectra of the N=2 superlatrice was measured to validate the measured optical constants. The N=2 superlattice grown at 600 C was transferred from the sapphire growth substrate onto a 100 nm template-stripped gold mirror using a potassium hydroxide (KOH) solution-based wet transfer process. We additionally calculated the reflectance spectra using a transfer matrix model (TMM). Figure S29 shows the measured (green) and calculated (blue and orange) normal incidence reflectance spectra of the N = 2 superlattice on the Au mirror. The measured data is normalized to the measured reflectance spectrum from the bare Au mirror, while the calculated spectra are normalized to a TMM spectrum for the bare Au.

The blue line is the TMM reflectance spectrum for the N = 2 superlattice, treating the superlattice as a single layer with the complex refractive indices shown in Fig S30. The reflectance spectra are normalized to the calculated spectra of a 100 nm Au mirror using the same model. The calculated N=2 superlattice spectrum fits the experimental data relatively well. The differences between the experimental and calculated superlattice spectra may be due to surface roughness, strain, or impurities generated during the wet transfer process. The orange line is a transfer matrix model (TMM) simulation of individual layers of $MoSe_2$ and $WSe_2$ stacked together, using the measured complex refractive indices of the individual layers. It is obvious that this does not describe the optical phenomena we see.

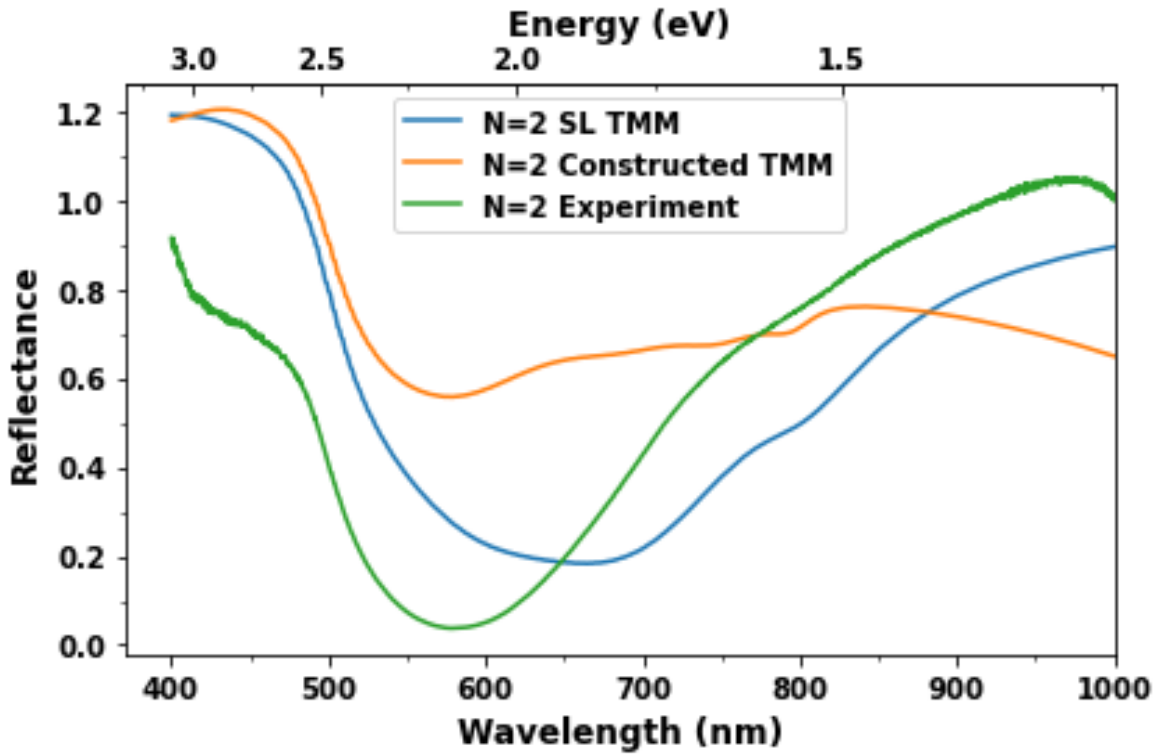

*Figure S29: Measured (solid black line) and calculated (dashed lines) normal incidence reflectance spectra of an N = 2 superlattice on a 100 nm Au mirror. The black dashed line is a transfer matrix model of the system using the measured complex refractive index for N = 2. The red dashed line is a transfer matrix model of a system constructed from alternating layers of $MoSe_2$ and $WSe_2$ using the measured refractive indices of the individual layers. The measured spectrum is normalized to the measured bare Au mirror spectrum; the calculated spectra are normalized to a calculated Au spectrum.*

Comparing the TMM results to the measured reflectance results for the N = 2 superlattice validates the measured optical constants and – by comparing the measured reflectance spectra to the TMM results for a simple N = 2 stack of WSe$_2$ and MoSe$_2$ layers with individual optical constants – further supports the theory that the change in the optical constants of the superlattice is due to the novel electronic structure of superlattice and not simply explained by the optical effects of the sum of its parts.